\documentclass[fleqn,usenatbib]{mnras}
\usepackage{fix-cm}

\usepackage{newtxtext,newtxmath}

\usepackage[T1]{fontenc}

\DeclareRobustCommand{\VAN}[3]{#2}
\let\VANthebibliography\thebibliography
\def\thebibliography{\DeclareRobustCommand{\VAN}[3]{##3}\VANthebibliography}

\usepackage{graphicx}	
\usepackage{amsmath}	

\usepackage{hyperref}
\usepackage{booktabs}
\usepackage{amsmath}
\usepackage{ulem}
\usepackage{footmisc} 
\usepackage{color}

\title[NGC 7789]{A Binary-Based Reassessment of the Age and Stellar Properties of NGC~7789 Using Twelve Binary Components}

\author[K. Yakut et al.]{
Kadri Yakut,$^{1,2}$\thanks{E-mail: kadri.yakut@ege.edu.tr(KY)}
Belinda Kalomeni$^{1}$
and
Saul Rappaport$^{3}$
\\
$^{1}$Department of Astronomy and Space Sciences, Faculty of Science, Ege University, 35100, {\.I}zmir, Türkiye\\
$^{2}$Institute of Astronomy, The Observatories, Madingley Road, Cambridge CB3 OHA, UK\\
$^{3}$Department of Physics, Kavli Institute for Astrophysics and Space Research, M.I.T., Cambridge, USA\\
}
\date{Accepted XXX. Received YYY; in original form ZZZ}

\pubyear{\the\year{}}

\begin{document}
\label{firstpage}
\pagerange{\pageref{firstpage}--\pageref{lastpage}}
\maketitle


\begin{abstract}
We present a binary-based reassessment of the age of the intermediate-age open cluster NGC\,7789, together with well-constrained stellar parameters for twelve components in six SB2 systems, including two eclipsing binaries. Our analysis employs a unified modelling framework that combines radial-velocity orbits, TESS light curves, and blue-to-IR spectral energy distributions (SEDs), providing a robust alternative to traditional isochrone-based age determinations. By adopting common cluster-wide parameters (age, distance, and line-of-sight extinction) when solving for the stellar parameters of the binary components, we obtain a coherent set of masses, radii, effective temperatures, and luminosities for all twelve stars. The combined SED, eclipsing-binary, and radial-velocity analysis yields a well-constrained cluster age of $1.26 \pm 0.09$\,Gyr and an extinction of $A_V = 0.90 \pm 0.05$\,mag, while remaining consistent with the \textit{Gaia} DR3 distance of $d \simeq 2.06$\,kpc used as an external prior. An independent Gaia\,DR3 astrometric analysis gives a distance of $2082 \pm 142$\,pc and confirms the membership of all six systems. The twelve binary components occupy the turnoff and subgiant regions of the cluster, enabling stringent evolutionary tests: in the radius--mass, radius--temperature, and temperature--mass diagrams, they show excellent agreement with modern stellar evolution models for the derived cluster parameters.
NGC\,7789 has long been affected by age discrepancies arising from MSTO broadening, rotation, and differential reddening, with literature estimates spanning $1.1$ to $1.6$\,Gyr. Our binary-anchored analysis significantly narrows this range and demonstrates that a self-consistent set of parameters emerges when dynamical SB2 information, eclipsing-binary geometry, and SED constraints are jointly analyzed. NGC\,7789 thus serves as a valuable benchmark for multi-observable, binary-based age determinations in open cluster studies.
\end{abstract}




\section{Introduction}
\label{sec:intro}

Open clusters serve as fundamental laboratories for testing stellar evolution theory, calibrating age–activity and age–rotation relations, and tracing the chemical and dynamical history of the Galactic disk. Their ages are therefore of central importance, yet age determinations remain sensitive to the adopted methodology. Traditionally, the ages of open clusters have been inferred through (i) isochrone fitting to colour–magnitude diagrams, (ii) asteroseismic constraints, (iii) white-dwarf cooling sequences, (iv) gyrochronology, and (v) eclipsing-binary (EB) based age determinations, which exploit precisely measured stellar masses and radii. Each of these has its own advantages and systematic limitations. In two recent studies \citep{Yakut2025a, Yakut2025b}, we introduced a complementary, physically grounded approach in which the ages of clusters are derived by jointly modelling the spectral energy distributions (SEDs), radial velocities (RVs), and (when available) light curves (LCs) of multiple double-lined binary systems. Because this method relies on stellar masses, radii, and temperatures that are tightly constrained by binary dynamics and multi-observable modelling—rather than on CMD morphology alone—it provides an independent and sensitive probe of cluster ages. We first applied this technique to the old open cluster NGC~188 (see also the case of the very old open cluster NGC~6791; \citealt{Yakut2015}), obtaining an age of $6.41 \pm 0.33$~Gyr, and subsequently to the intermediate-age cluster NGC~2506, deriving $1.94 \pm 0.03$~Gyr. In the present work, we extend the same methodology to the younger cluster NGC~7789.

NGC~7789 ($\alpha_{\text{J2000}} = 23^{\rm h}57^{\rm m}19^{\rm s}$, $\delta_{\text{J2000}} = +56^{\circ}43^{\prime}28^{\prime\prime}$) is a rich, well-studied, intermediate-age open cluster located in the constellation Cassiopeia \citep{AnthonyTwarog2025, Friel2002, Nagarajan2023}. With Galactic coordinates $(l,b) \approx (115.5^{\circ}, -5.4^{\circ})$ \citep{Ferreira2025}, the cluster lies in the direction of the Galactic anticentre and serves as an important benchmark system for tracing stellar evolution and Galactic disk chemo-dynamics. NGC~7789 hosts a dense stellar population, including a well-developed main-sequence turnoff (MSTO), a prominent red giant branch, and an unusually large population of blue straggler stars (BSS) \citep{AnthonyTwarog2025, Friel2002}. Estimates based on deep Gaia photometry suggest that the cluster may contain as many as $\sim3000$ members down to $G \approx 21$ mag \citep{Gao2018}, making it one of the richer intermediate-age clusters in the solar neighbourhood.

Historically, the cluster was first studied photometrically by \citet{Burbidge1958} using $UBV$ photometry for nearly 700 stars, yielding an early reddening estimate of $E(B-V)=0.24$. Modern combined photometric and Gaia-based analyses have revised this value upward: \citet{AnthonyTwarog2025} reported $E(B-V)=0.30\pm0.02$, and \citet{Ferreira2025} derived $E(B-V)=0.328\pm0.008$ from Gaia DR3 data.
NGC~7789 is known to be affected by substantial interstellar reddening owing to its relatively low Galactic latitude, which has long complicated its photometric interpretation.

Chemical abundance studies indicate that NGC~7789 has an overall near-solar iron abundance.
Early photometric metallicity estimates suggested subsolar values, such as $[\mathrm{Fe}/\mathrm{H}] = -0.26\pm0.10$ \citep{Friel2002}, whereas modern high-resolution spectroscopic analyses consistently converge toward solar composition, including $[\mathrm{Fe}/\mathrm{H}] = -0.04\pm0.05$ \citep{Tautvaisiene2005} and
$[\mathrm{Fe}/\mathrm{H}] = +0.02\pm0.04$ \citep{Jacobson2011}.
At the same time, detailed chemical studies reveal that the cluster is not chemically simple, showing significant enhancements in neutron-capture elements—particularly barium—with $[\mathrm{Ba}/\mathrm{Fe}] \approx +0.47$ to $+0.48$ \citep{Overbeek2015,Pancino2010}, along with mild $\alpha$-element enrichment and evolutionary C–N–O abundance patterns
associated with dredge-up and extra mixing processes \citep[e.g.,][]{Overbeek2015,Nagarajan2023}.

Estimates of the cluster's age vary across the literature. Traditional photometric analyses based on colour--magnitude diagram fitting have typically yielded ages in the range $1.4$–$1.6$~Gyr
\citep[e.g.][]{Gim1998, Ferreira2025}. Using \textit{Gaia} DR2 astrometry and homogeneous isochrone fitting,
\citet{CantatGaudin2020} derived an age of $\sim1.2$~Gyr for NGC~7789, highlighting the sensitivity of the inferred age to the adopted methodology and treatment of extinction. Evidence has also been presented for an extended main--sequence turnoff (eMSTO): \citet{Ovelar2020} interpreted the observed MSTO width as corresponding to an age spread of $\sim390$~Myr, suggesting a representative age of $1070 \pm 390$~Myr, although the interpretation remains debated and may be linked to stellar rotation, overshooting, or differential reddening \citep{AnthonyTwarog2025}.
Overall, the dispersion in published age estimates primarily reflects the combined effects of differential reddening, high binary fraction, and rotation--driven broadening of the MSTO. For clarity, a consolidated summary of previously published age, distance, metallicity, and reddening estimates for NGC~7789 is provided in Table~\ref{tab:lit}.

Radial-velocity surveys have played a central role in establishing cluster membership and identifying binary systems. The WIYN Open Cluster Study (WOCS) survey identified 624 likely members and derived orbits for 81 spectroscopic binaries with orbital periods ranging from $1.45$ to $4200$ days \citep{Nine2020}. From these surveys we selected six SB2 systems with secure Gaia DR3 membership, well-determined RV curves, and reliable broad-band SED coverage. The systemic radial velocity of the cluster lies between $-54$ and $-58$ km\,s$^{-1}$, with a consensus value of approximately $-54.5$ km\,s$^{-1}$. 

Binary evolution is known to play a fundamental role in shaping the observed properties of star clusters, affecting both their colour--magnitude morphologies and the interpretation of cluster ages. 
Theoretical and population--synthesis studies have long shown that mass transfer, angular momentum loss, and mergers in binary systems can substantially alter stellar evolutionary paths, producing rejuvenated stars and blue straggler populations that deviate from single--star isochrones \citep[e.g.,][]{Hurley2001,Yakut2009}. 
Observationally, blue straggler stars are therefore widely regarded as indirect tracers of past and ongoing binary interaction in clusters \citep[e.g.,][]{Ferraro2009,Perets2009}. 
In NGC~7789, ultraviolet observations have revealed hot companions and extremely low--mass white dwarfs in a significant fraction of blue stragglers, strongly supporting mass--transfer formation channels \citep{Vaidya2022,Vaidya2024}.

\begin{table*}
\centering
\caption{Summary of derived parameters for the NGC~7789 open cluster from past and recent investigations.}
\label{tab:lit}
\begin{tabular}{lcccl}
\hline
Age & Distance & [Fe/H] & $E(B-V)$ & Reference \\
Gyr & kpc & dex & mag &   \\
\hline
$1.1$           & $1.86$            &                  & 0.35            & \cite{Mazzei1988} \\
$1.2$           & $2.45\pm 0.15$    &                  & $0.33 \pm  0.03$& \cite{Martinez1994} \\
                &                   & $-0.04\pm0.05$   &                 & \cite{Tautvaivsiene2005} \\
$1.2$           & $2.08\pm 0.01$    & $+0.023$         &                 & \cite{CantatGaudin2018} \\
1.4             & 1.80              &                  & 0.28            & \cite{Gao2018} \\
$1.61\pm 0.05$  &$1.91 \pm 0.03$    & $+0.026$         & $0.31\pm0.02$   & \cite{Dias2021} \\   
$1.2$           & $1.8$             & $+0.023$         &     0.27        & \cite{Nine2020} \\
1.07            &                   &                  &                 & \cite{Ovelar2020}\\
                &                   & $-0.013\pm0.023$ &                 & \cite{Spina2021}\\
$1.42 \pm 0.12$ & $1.92 \pm 0.08$   &                  & $0.28$          & \cite{Vaidya2022} \\
                &                   & $-0.02 \pm 0.05$ &                 & \cite{Nagarajan2023} \\
                &                   & $0.00\pm0.07$    &                 & \cite{CarbajoHijarrubia2024}\\                
$1.60 \pm 0.10$ & $1.90$            & $0.04 \pm 0.05$  &                 & \cite{Balan2024} \\
                &                   & $-0.01 \pm 0.04$ &                 & \cite{Nine2024} \\
$1.51 \pm 0.20$ & $1.91 \pm 0.03$   & $+0.05 \pm 0.07$ &$0.328 \pm 0.008$& \cite{Ferreira2025} \\
\hline
$1.34 \pm 0.18$ & $1.96 \pm 0.21$ & $+0.01 \pm 0.03$ & $0.31 \pm 0.03$ & Mean $\pm$ std.\,dev. \\
\hline
   $1.26 \pm 0.09$   & ... & $\gtrsim +0.1$         &    ${0.29 \pm 0.02}$    & This study \\
\hline
\end{tabular}
\end{table*}

A consolidated overview of the literature values for age, metallicity, reddening, and distance is provided in Table~\ref{tab:lit}, which summarizes results from both classical and modern investigations.

Modern Gaia-based analyses provide refined estimates of the distance and reddening of the cluster. Gaia-based studies place the cluster at a distance of $\sim2.1$ kpc \citep[e.g.][]{CantatGaudin2018, Nine2020}, consistent with more recent DR3 analyses. Similarly, reddening estimates put the cluster in the range $0.27$–$0.33$ mag, consistent with recent high-quality photometry and spectroscopy.

In two recent studies of NGC\,188 and NGC\,2506, we demonstrated that combining spectroscopic orbital solutions with broad-band spectral energy distributions of multiple binary systems provides a powerful and physically grounded alternative to traditional isochrone fitting \citep{Yakut2025a, Yakut2025b}. This joint RV+LC+SED methodology yields self-consistent cluster ages, distances, and extinction values by using directly measured stellar masses and radii rather than inferring them from photometry alone. In the present study of NGC\,7789, by anchoring the cluster parameters to six binaries comprising twelve well-characterised stellar components, we obtain one of the most precise and dynamically validated determinations of the cluster age to date.

In the present study, we apply the same methodology to NGC~7789 using six double-lined spectroscopic binaries, including two eclipsing systems for which joint light-curve and radial-velocity solutions can be obtained. Section~2 describes the observational material used in this work. Section~3 presents the radial-velocity and light-curve modelling. Section~4 introduces the SED analysis and joint fitting framework. Section~5 summarizes the Gaia-based astrometric validation. Section~6 discusses the combined results and their implications for the age, distance, reddening, and chemical composition of NGC~7789.


\section{Observations}
\label{sec:obs}

The observational data used in this study consist of three complementary components: 
(1) TESS photometry, 
(2) ground–based spectroscopic radial–velocity (RV) measurements, and 
(3) broadband spectral energy distributions assembled from archival surveys.  
These datasets provide independent and mutually constraining diagnostics on the physical properties of the six binary systems analysed in this work. 
Basic astrometric and photometric information for all six double--lined spectroscopic binaries considered in this study is summarised in Table~\ref{tab:basic_parameters_all}.

\subsection{Radial–Velocity Measurements}

\begin{figure*}
\centering
\includegraphics[width=0.3\linewidth]{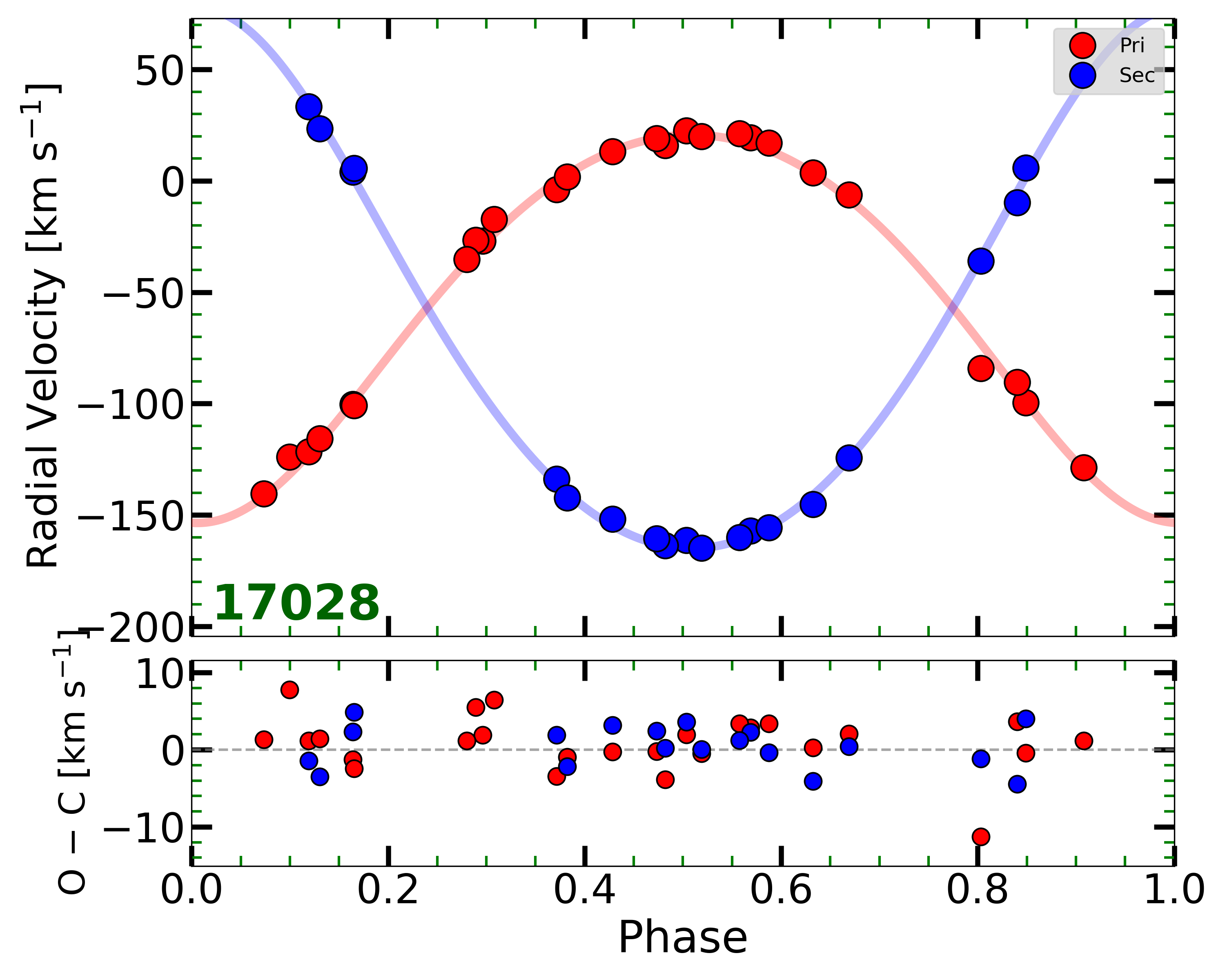}
\includegraphics[width=0.3\linewidth]{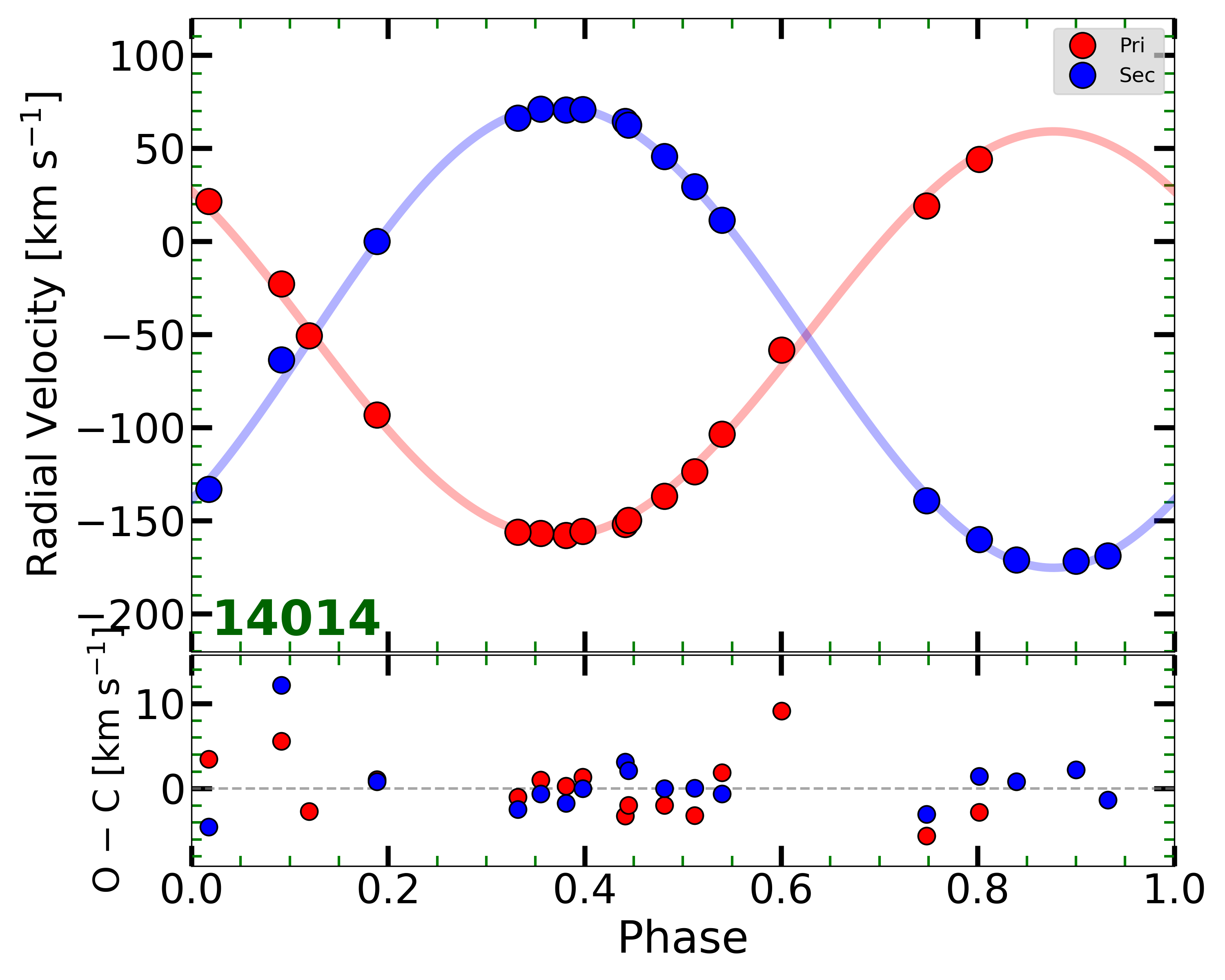}
\includegraphics[width=0.3\linewidth]{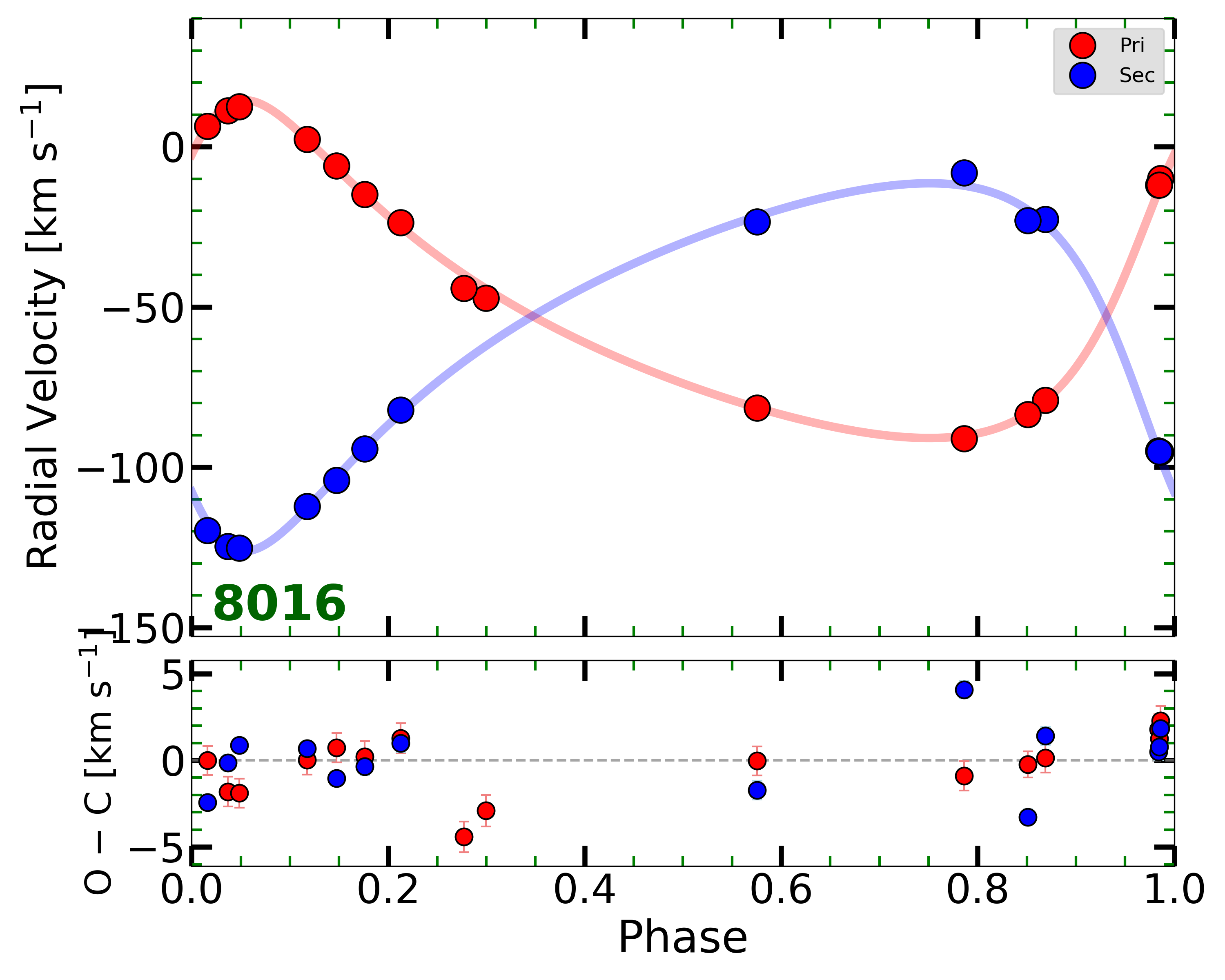}\\
\includegraphics[width=0.3\linewidth]{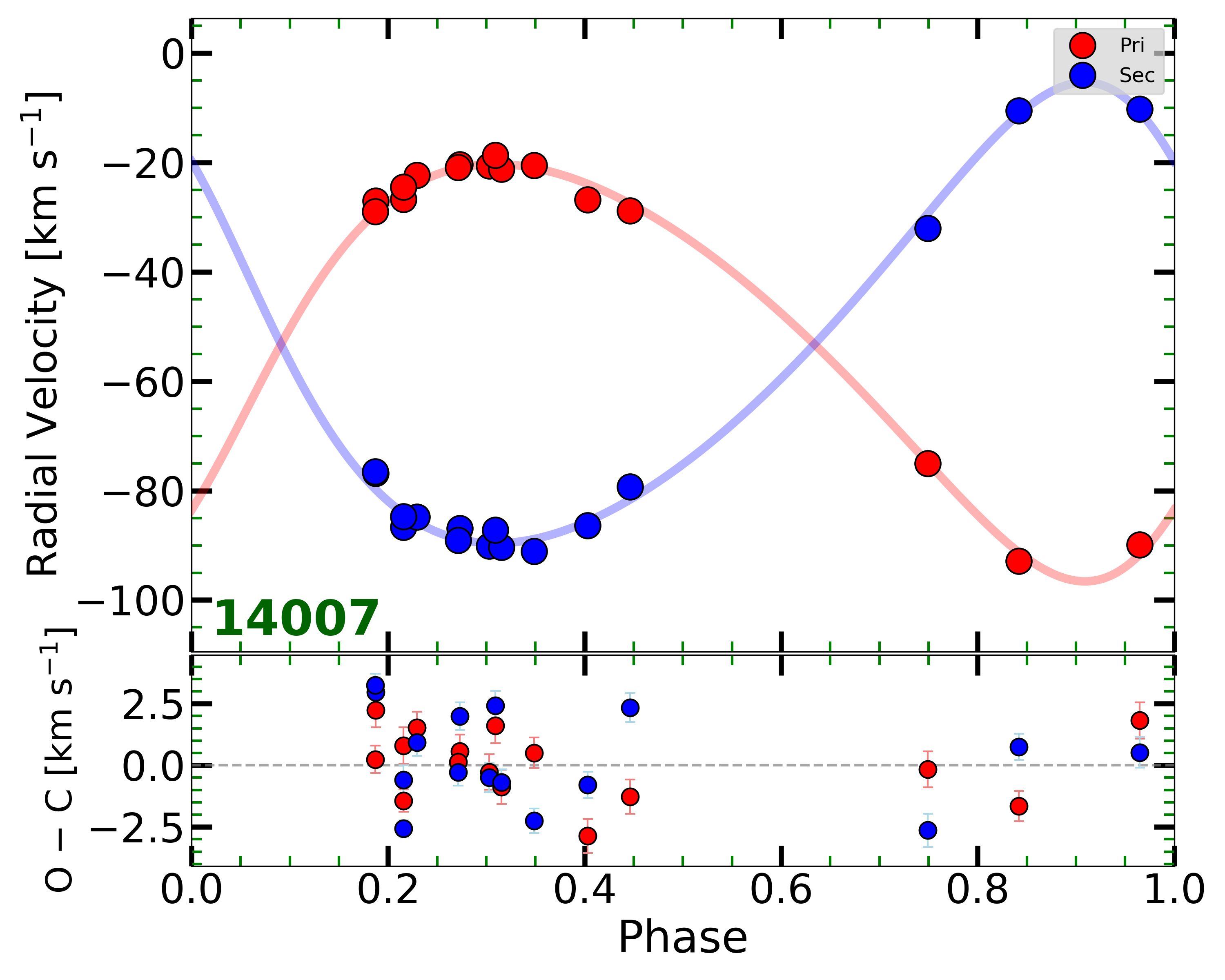}
\includegraphics[width=0.3\linewidth]{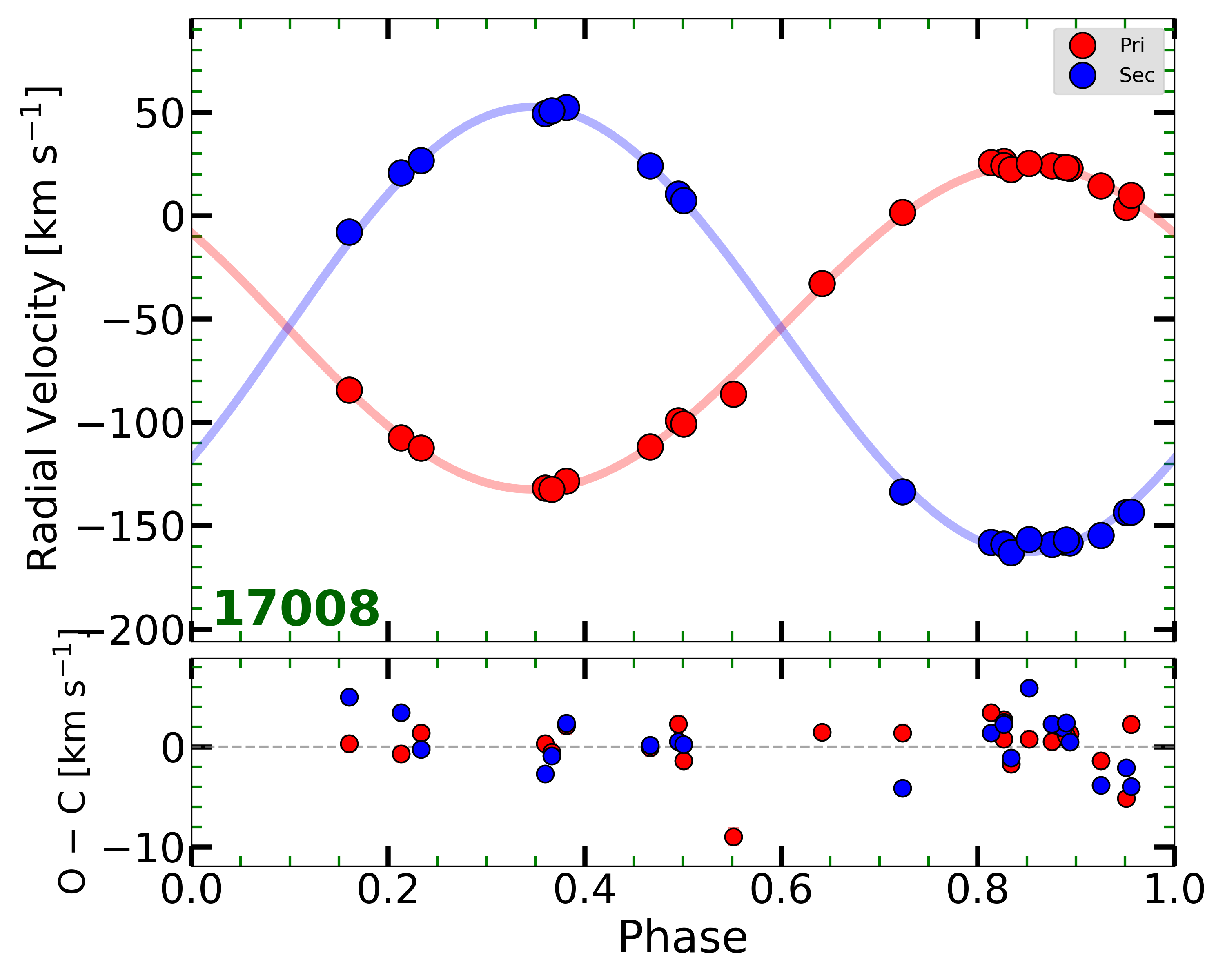}
\includegraphics[width=0.3\linewidth]{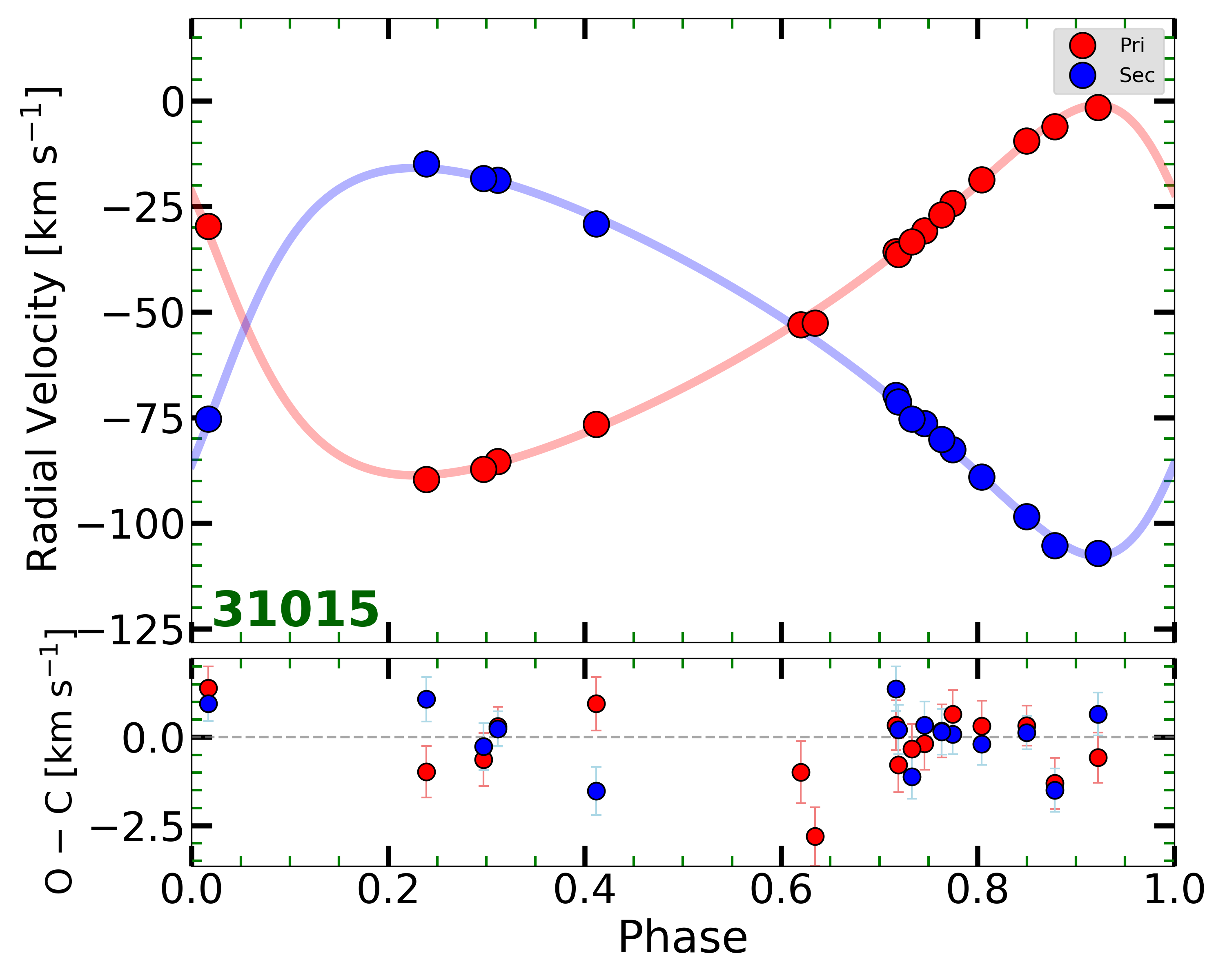}
\caption{Observed (circles) and modelled (lines) radial velocities for the six selected binary systems in NGC~7789. Residuals (O-C) are shown below each panel. Red color denote the primary component, blue for the secondary.}
\label{fig:RVs}
\end{figure*}

High–resolution spectroscopy for all six binary systems was obtained by \citet{Nine2020}, 
who provided double–lined RV measurements for both the primary and secondary components.  
These data form the basis of our orbital solutions (Sect.~\ref{sec:rvlc}).  
We adopted the heliocentric Julian dates (HJDs), RVs, and published measurement weights (\texttt{CH} values), 
but performed our fits using both weighted and unweighted approaches to ensure robustness.  
The phase–folded RV curves and the best–fitting orbital models are displayed in Figure~\ref{fig:RVs}.  
The derived orbital parameters for the six systems are summarised in Table~\ref{tab:rvorbit}.

\subsection{TESS Photometry}

NGC\,7789 was observed by the \textit{Transiting Exoplanet Survey Satellite} (TESS) during Sectors~17 and~24. 
We extracted light curves for all six target systems using the publicly available TASOC and SPOC pipelines, 
and subsequently performed detrending and outlier removal following the procedures described in our earlier cluster studies.  
Among the six systems, only two --- WOCS\,17028 and WOCS\,14014 --- exhibit detectable eclipses in the TESS photometry.  
For these two binaries, we carried out a full eclipsing–binary analysis (Sect.~\ref{sec:rvlc}), 
while for the remaining four systems the TESS fluxes were used solely for variability checks and contamination assessment.
The phase–folded TESS light curves for the two eclipsing systems are shown in Figure~\ref{fig:LCs}, 
where the insets highlight the eclipse minima.

\subsection{Spectral Energy Distributions}

We constructed broadband SEDs for all six binaries using archival photometry compiled through the VizieR database \citep{ochsenbein00}. The dataset spans the ultraviolet to the infrared, drawing on measurements from \textit{GALEX} \citep{bianchi17}, APASS \citep{APASS2016}, Pan-STARRS \citep{chambers16}, 2MASS \citep{2MASS}, and \textit{WISE} \citep{WISE}. For each system, the photometry was assembled using \textit{Gaia} DR3 source identifiers to ensure a consistent and unambiguous association of flux measurements across all wavelength bands. The final SEDs, shown in Figure~\ref{fig:seds}, exhibit smooth flux distributions with no significant discontinuities between adjacent passbands, except perhaps at the Balmer discontinuity near 3645\,\AA.
These SEDs form one of the three pillars of our joint SED+EB+RV modelling approach (Sect.~\ref{sec:SEDs}), 
providing strong constraints on stellar luminosities, radii, and effective temperatures.

Together, the TESS photometry, high–precision RV orbits, and broadband SEDs constitute a complete and complementary observational framework.  
This multi–observable dataset enables the simultaneous determination of the stellar and orbital properties of the six binaries, 
while also providing the foundation for the global cluster–parameter inference presented in the subsequent sections.

\begin{table*}
\scriptsize
\caption{Astrometric and photometric properties of the six double-lined binary systems analyzed in this study, including Gaia DR3 and TESS IDs, proper motions, parallaxes, and Gaia colours.}
\label{tab:basic_parameters_all}
\begin{tabular}{l r r r r r  r}
\hline
                                & WOCS 17028          &  WOCS 14014        & WOCS 8016          &  WOCS 14007         & WOCS 17008         & WOCS 31015           \\
                                & A                   & B                  &   C                &    D                &     E              &     F               \\
\hline
TIC ID                          & 417031808           & 65558198           & 416814228          & 416814233           & 416814240           & 65558397           \\
Gaia DR3 ID                     & 1995020800153469568 & 1995061104128742272& 1995066086290711424& 1995010835829510144 &1995010938908725376& 1995013000492937984 \\
Alias                           & NGC~7789 XZD 19     & V874 Cas           & NGC~7789 521       & UCAC4 734-105175    &J235747.61+564142.3 &                     \\
$\alpha$ ($^{\rm o}$)           & 359.683             & 359.134            & 359.300            & 359.419             & 359.448            & 359.132             \\
$\delta$ ($^{\circ}$)           & +56.851             & +56.738            & +56.853            & +56.689             & +56.695            & +56.675             \\
$\mu_{\alpha}$ (mas~yr$^{-1}$)  & -0.815              & -0.672             & -0.945             & -0.721              & -0.699             & -0.698              \\
$\mu_{\delta}$ (mas~yr$^{-1}$)  & -2.016              & -1.916             & -1.827             & -2.009              & -1.898             & -1.786               \\
$\varpi$ (mas)                  & 0.4697              & 0.4795             & 0.4777             & 0.4495              & 0.4728             & 0.4840              \\
$G$ (mag)                         & 13.769              & 13.828             & 13.320             & 13.719              & 13.427             & 14.572              \\
$G_{\rm{BP}}-G_{\rm{RP}}$ (mag) & 0.925               & 0.905              & 0.982              & 0.854               & 0.955              & 0.931               \\
\hline

\end{tabular}
\end{table*}

\section{Light and Radial Velocity Curve Modeling}
\label{sec:rvlc}

The spectroscopic orbital solutions for all six SB2 systems were derived from the complete set of radial‐velocity (RV) measurements published by \citet{Nine2020}.  
Since the original RV files list a third-column quantity (“CH”) rather than formal measurement uncertainties, we adopted two complementary strategies.  
Our primary analysis employed strictly unweighted RV fits in order to avoid propagating unknown or non-standard error estimates into the orbital solution.  
A secondary set of tests was performed in which the radial--velocity uncertainties were uniformly increased to assess the sensitivity of the fitted semi--amplitudes to reasonable changes in the assumed error scale.
For each binary, an initial least-squares solution was followed by a full Markov Chain Monte Carlo (MCMC) exploration of the posterior distributions.  
The resulting orbital elements, including ($K_1$, $K_2$), systemic velocity, and eccentricity when significant, are listed in Table~\ref{tab:rvorbit}.

\begin{figure}
\centering
\includegraphics[width=0.95\linewidth]{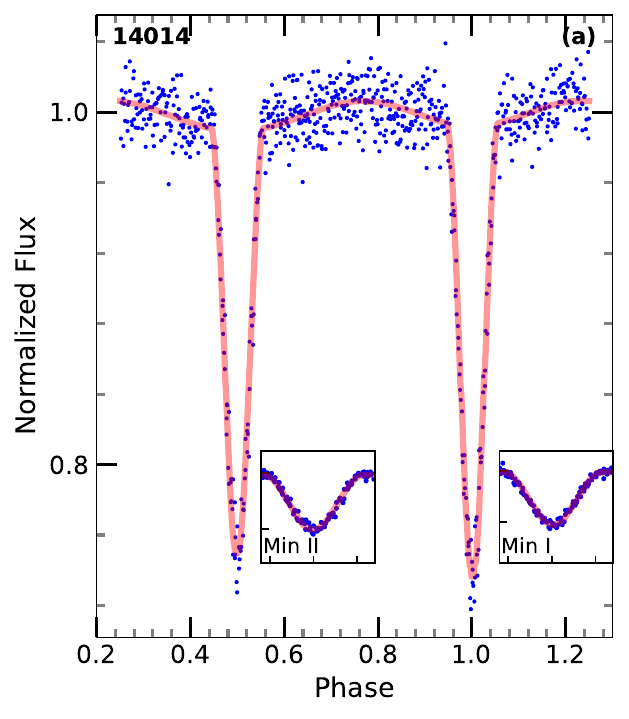}\\
\includegraphics[width=0.95\linewidth]{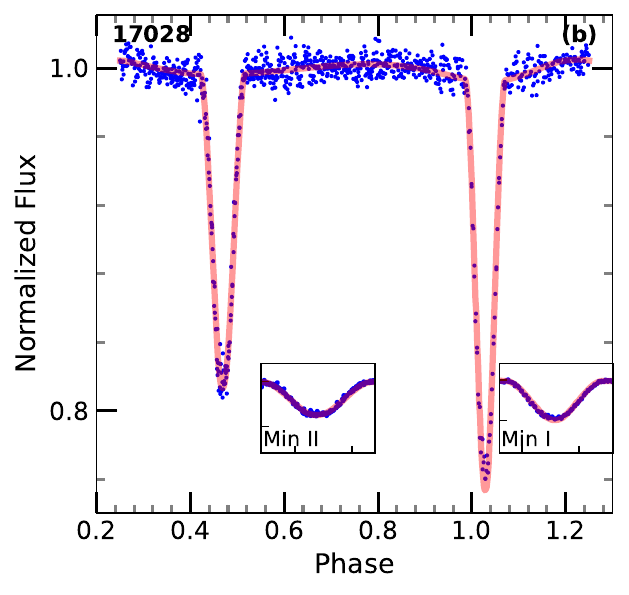}
\caption{The observed and the modelled (solid lines) light curves of WOCS 14014 and WOCS 17028; insets provide zoomed views of the primary (Min I) and secondary (Min II) minima to illustrate the goodness of fit.} \label{fig:LCs}
\end{figure}

Among the six systems, only WOCS\,14014 and WOCS\,17028 exhibit eclipses in the available TESS photometry.  
For these two sources, we performed a joint LC+RV solution using the MCMC-refined RV orbits as dynamical anchors.  
This combined modelling yields purely geometric constraints on the absolute masses and radii, independent of stellar evolutionary calculations, and provides a robust foundation for the subsequent SED-based inference.  
For the remaining four systems, only RV constraints were used, and their masses remain tied to the SED+evolution-track solution discussed in Sect.~\ref{sec:SEDs}. The modelled lightcurves for WOCS 14014 and 17028 are compared to the TESS data in Fig.~\ref{fig:LCs}. We also tested solutions in which a third-light contribution ($\ell_3$) was treated as a free parameter. In both systems, the fits converged to negligible or non-physical values ($\ell_3 \le 0$ for WOCS\,17028 and $\ell_3 \simeq 0$ for WOCS\,14014), indicating that no detectable photometric third-light contribution is supported by the data.

In summary, the radial--velocity data provide the primary constraints on the orbital semi--amplitudes ($K_1$, $K_2$) and mass ratios for all six systems, thereby defining the dynamical framework of each binary. The RV data plus the eclipses in two of the binaries lead directly to the determination of the masses and radii for four of the stars.  For the other eight stars, the RV data primarily fix the mass ratios in each of these four binaries, and the SED analysis (see next section) is able to utilize this information to determine the masses for all twelve stars.

\begin{table*}
\centering
\caption{Orbital parameters of the six double-lined spectroscopic binary systems analyzed in this study. For each system, the following parameters are listed: $K_1$ and $K_2$ are the radial velocity semi-amplitudes of the primary and secondary components, $P$ is the orbital period, $e$ is the orbital eccentricity, $\omega$ is the argument of periastron, T$_{\rm 0}$ is the time of periastron passage, and $\gamma$ is the systemic velocity}
\label{tab:rvorbit}
\resizebox{\textwidth}{!}{%
\begin{tabular}{llllllll}
\toprule
ID & $T_0$ & $P_{\text{orb}}$ & $e$ & $\omega$ & $V_{\gamma}$ & $K_1$ & $K_2$ \\
 & (HJD - 2400000) & (days) & & (deg) & (km s$^{-1}$) & (km s$^{-1}$) & (km s$^{-1}$) \\
\midrule
17028 & $54711.7634 \pm 0.0107$ & $3.55387 \pm 0.00001$ & $0.115 \pm 0.004$ & $173 \pm 1$ & $-56.28 \pm 0.16$ & $87.41 \pm 0.37$ & $121.41 \pm 0.55$ \\
14014 & $55371.3462 \pm 0.0645$ & $2.36602 \pm 0.00000$ & $0.013 \pm 0.002$ & $57 \pm 10$ & $-50.11 \pm 0.20$ & $109.20 \pm 0.42$ & $123.86 \pm 0.31$ \\
8016 & $55313.3398 \pm 0.0428$ & $30.97954 \pm 0.00086$ & $0.409 \pm 0.005$ & $312 \pm 1$ & $-52.96 \pm 0.22$ & $52.67 \pm 0.36$ & $57.22 \pm 0.36$ \\
14007 & $54571.7490 \pm 0.1365$ & $23.16631 \pm 0.00120$ & $0.224 \pm 0.016$ & $228 \pm 2$ & $-53.13 \pm 0.22$ & $38.64 \pm 0.70$ & $42.40 \pm 0.75$ \\
17008 & $55011.4102 \pm 0.0746$ & $2.76074 \pm 0.00000$ & $0.009 \pm 0.001$ & $43 \pm 10$ & $-54.18 \pm 0.15$ & $78.72 \pm 0.25$ & $107.38 \pm 0.26$ \\
31015 & $55558.2274 \pm 0.1401$ & $34.41091 \pm 0.00553$ & $0.356 \pm 0.010$ & $58 \pm 2$ & $-53.26 \pm 0.19$ & $43.81 \pm 0.50$ & $45.82 \pm 0.51$ \\
\bottomrule
\end{tabular}%
} 
\end{table*}

\begin{table}
\begin{center}
\caption{Obtained fundamental parameters for WOCS 14014 and WOCS 17028 based on simultaneous LC and RV solutions. 
The standard errors 1$\sigma$ in the last digit are given in parentheses.} \label{tab:LCRV_results}
\begin{tabular}{lllllll}
\hline
Parameter                                              &  14014           &  17028          \\
\hline
Initial epoch, T$_{\rm 0}$   (day)                     & 58960.8043(97)   & 58961.3848(74)  \\
Period, P  (day)                                       & 2.3660116(4)     & 3.5540157(5)  \\
Geometric parameters:                                  &                  &                \\
Inclination, i ${({^\circ})}$                          & 83.1(2)          & 82.3(2)        \\
Eccentricity, e                                        & 0.011(3)         & 0.110(10)       \\
$\Omega _{1}$                                          & 5.535(29)        & 6.385(22)        \\
$\Omega _{2}$                                          & 7.456(46)        & 7.079(29)       \\
Fractional radii of pri.                               &                  &                \\ 
$R_1/a$                                                & 0.2185(13)       & 0.1803(7)       \\
Fractional radii of sec.                               &                  &                \\ 
$R_2/a$                                                & 0.1390(11)       & 0.1245(6)       \\
Radiative parameters:                                  &                  &                \\ 
T$_{\rm eff, \,1}$  (K)                                & 6917(123)        & 6727(161)       \\
T$_{\rm eff, \,2}$  (K)                                & 6824(121)        & 6442(156)       \\
Light ratios                                           &                  &                \\
$(\frac{l_1}{l_1+l_2})_{\rm TESS}$(\%)                 & 73               & 71             \\
\hline
Mass, $M_{1}$ ($\rm{M_{\odot}}$)                       & 1.686(34)        & 1.963(23)      \\
Radius, $R_{1}$  ($\rm{R{\odot}}$)                     & 2.398(43)        & 2.651(55)     \\
Luminosity, $L_1$ ($\rm{L_{\odot}})$                   & 11.8(1.2)        & 12.9(1.7)      \\
Surface gravity, $\log g_1$ (cgs)                      & 3.906            & 3.885           \\
Mass, $M_{2}$ ($\rm{M_{\odot}}$)                       & 1.487(30)        & 1.414(17)       \\
Radius, $R_{2}$  ($\rm{R_{\odot}}$)                    & 1.526(27)        & 1.830(38)        \\
Luminosity, $L_{2}$ ($\rm{L_{\odot}})$                 & 4.5(5)           & 5.2(0.7)          \\
Surface gravity, $\log g_{2}$ (cgs)                    & 4.244            & 4.064              \\
\hline 
\end{tabular}
\end{center}
\end{table}

\section{Joint Sed Fitting for the Six Binary SEDs}
\label{sec:SEDs}

Before describing the individual SED fits, we briefly summarise the structure of the joint modelling;
see also \citet{Yakut2025a,Yakut2025b} for further details.
The analysis simultaneously solves for the parameters of all twelve stars together with a common
set of cluster properties.
The free parameters are: the masses of the twelve stars, and a global cluster age, distance (if the
prior is opened up), extinction, and metallicity (usually left fixed), as well as the orbital
inclinations of the four binaries without eclipses. This makes a total of between 18 and 20 free
parameters to be determined.
We utilise a Markov Chain Monte Carlo (MCMC) code to find the best fit to these parameters and their
uncertainties. 

The observational constraints comprise the following items: 12 RV semi-amplitudes ($K_1$ and $K_2$ for each of the six binaries), their orbital periods and eccentricities (taken to have negligible uncertainties), the orbital inclination angles for two of the six binaries, and approximately 150 broadband SED flux measurements spanning the blue to the mid-infrared.

The SED fitting code also utilizes stellar evolution constraints provided by MIST evolution tracks \citep{Choi2016,Dotter2016,Paxton2011,Paxton2015,Paxton2019} which link stellar mass, age, and metallicity to the stellar radii and effective temperatures. The synthetic spectral energy distributions are generated using the \citet{castelli03} atmosphere models. In the present analysis, we adopted atmosphere grids computed at a fixed $\log g = 4.0$ for all components. This choice is well justified, as the inferred surface gravities of our 12 stars stars have a mean $\log g$ of $3.99$ and an rms deviation from this value of 0.21 dex, implying a rather negligible impact on the derived SED-based parameters.

For each system, we compiled broadband photometry from the blue end of the spectrum to the mid-infrared (typically $0.35\,\mu$\,m $\lesssim \lambda \lesssim 12\,\mu$\,m) using the VizieR SED interface \citep{ochsenbein00}. The data set includes fluxes from large-scale surveys such as GALEX \citep{bianchi17}, Pan-STARRS \citep{chambers16}, SDSS \citep{gunn98}, 2MASS \citep{2MASS}, WISE \citep{WISE}, and \textit{Gaia} \citep{GaiaCollaboration2023}, supplemented where available by additional published optical photometry from the literature. All measurements were retrieved from VizieR in a homogeneous absolute flux scale, with standard zero-point calibrations already applied. The photometry was then de-reddened using the current MCMC trial value of the cluster extinction.  Because the photometry was taken non-simultaneously, and is subject to additional systematic effects (e.g. crowding, variability, and filter-transformation uncertainties), we imposed a minimum relative error floor of 5 per cent on all flux points, and adopted larger uncertainties for the most problematic bands (in particular the bluest point at 0.35 $\mu$m and the longest-wavelength WISE measurements). We adopted a minimum uncertainty floor of 5 per cent because the formal catalogue uncertainties are typically at the sub-percent level and therefore do not capture the true level of systematic effects affecting multi-survey SEDs. Since the quoted catalogue uncertainties are much smaller than 5 per cent, it makes no practical difference whether they are added in quadrature or replaced by the adopted floor. Aside from the eclipse phases in the two eclipsing systems, no evidence is found for intrinsic broadband variability at a level that would require a larger error floor than the adopted 5 per cent.

The line-of-sight extinction $A_V$ is treated as a single cluster-wide free parameter and is determined entirely from the joint SED fitting, rather than from the astrometric analysis. 
The adopted prior range on $A_V$ (0.74-0.98) is motivated by previous photometric studies of NGC~7789 and serves only as a weak, physically informed constraint. As a robustness check, we repeated the SED analysis with a substantially widened range on the $A_V$ prior (0.38-1.5). In this case, the posterior distribution shifts only slightly toward higher extinction values, while the inferred cluster age and stellar parameters remain unchanged within the quoted uncertainties. 
Given that unrealistically large extinctions are excluded by the literature, we therefore retain the adopted prior range in our final solution.

Photometric measurements at $\lambda \simeq 0.23\,\mu$m are available for four of the six binaries analysed in this study; these points were excluded from the SED fitting for the following reason. 
At such short wavelengths, the interstellar extinction corrections along the line of sight to NGC~7789 reach nearly an order of magnitude, rendering the de-reddened fluxes highly sensitive to uncertainties in the adopted extinction law (e.g. \citealt{Cardelli1989}). 
As a result, the propagated flux uncertainties exceed the level of precision required for the present analysis and would otherwise dominate the $\chi^2$ minimisation without providing commensurate physical constraints.

The fit to the SEDs proceeds roughly as follows.  At each link of the MCMC chain, the 12 stellar masses are chosen, as well  the age, distance, and exinction to the cluster.  The MIST evolution tracks are then used to evaluate the corresponding stellar radii and $T_{\rm eff}$.  These are used to compute the six model SED curves via the \citet{castelli03} stellar atmospheres.  The differences between the model and measured SEDs are recorded as contributions to $\chi^2$.  The masses, inclinations, and eccentricities are used to compute the 12 $K$ values. In turn, these are compared to the measured $K$ values and also contribute to $\chi^2$. The relative weighting between the contribution to $\chi^2$ from matching the $K$ values vs.~matching the SEDs was somewhat arbitrarily set at 80\%. However, the results are insensitive to whether this relative weighting lies anywhere between 50\% to 100\%. We ran many MCMC chains, with between 20 and 500 million links each.
To verify that the adopted relative weighting between the RV and SED contributions does not affect our conclusions, we performed additional tests in which the contribution of the RV semi-amplitudes to the total $\chi^2$ was varied over a wide range, from 17\% to 125\% of the SED contribution. Over this entire range, the inferred cluster age remains remarkably stable, with an rms scatter of only $\sim$1\%, and shows no systematic trend with the adopted weighting. While the absolute $\chi^2$ values naturally change as the RV contribution is up-weighted, reflecting known tensions for two systems, the derived cluster age and stellar parameters are essentially unchanged.

\begin{figure*}
\centering
\includegraphics[width=0.48\textwidth]{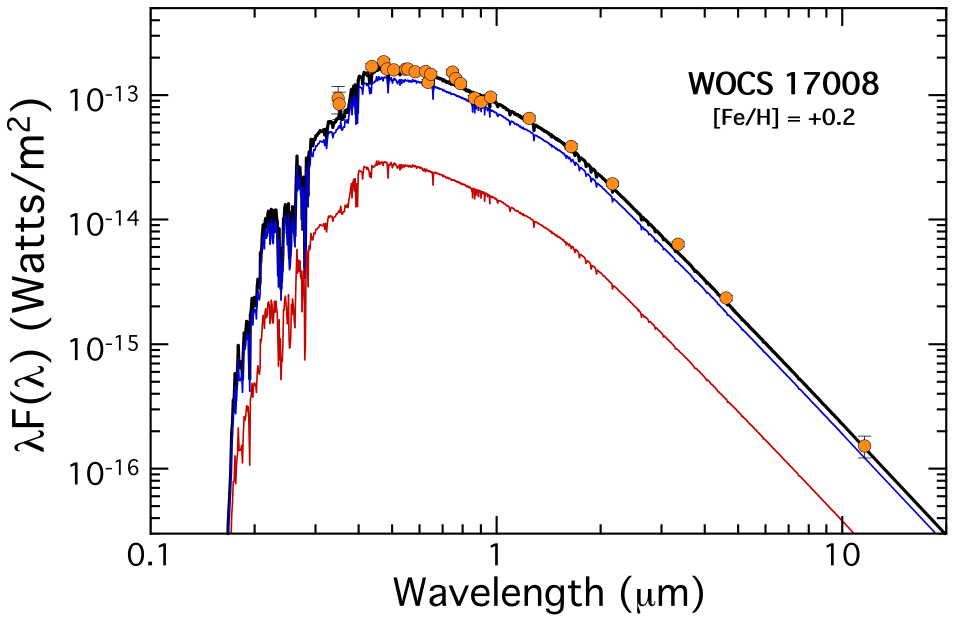}
\includegraphics[width=0.48\textwidth]{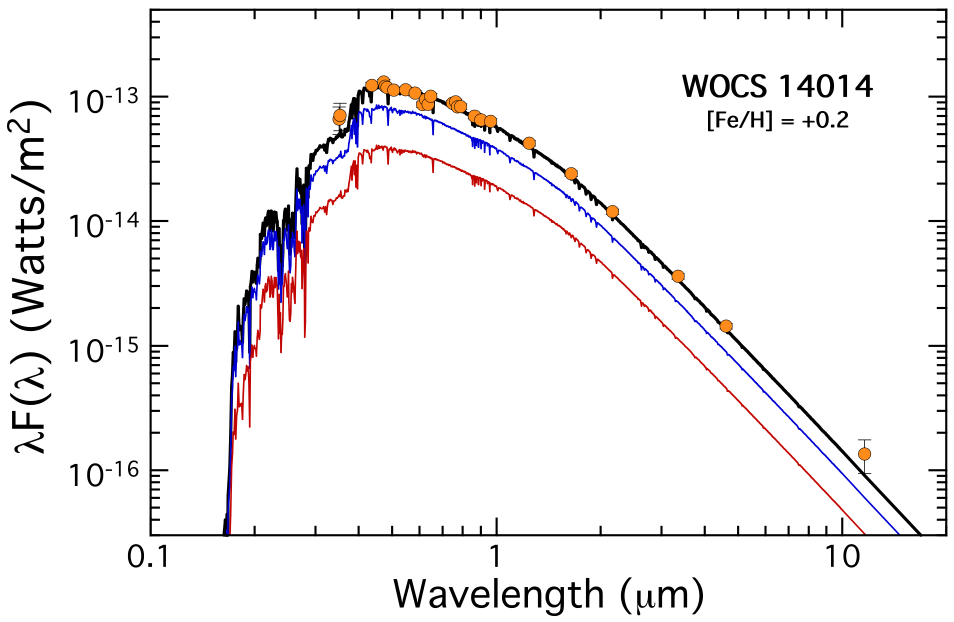}\\
\includegraphics[width=0.48\textwidth]{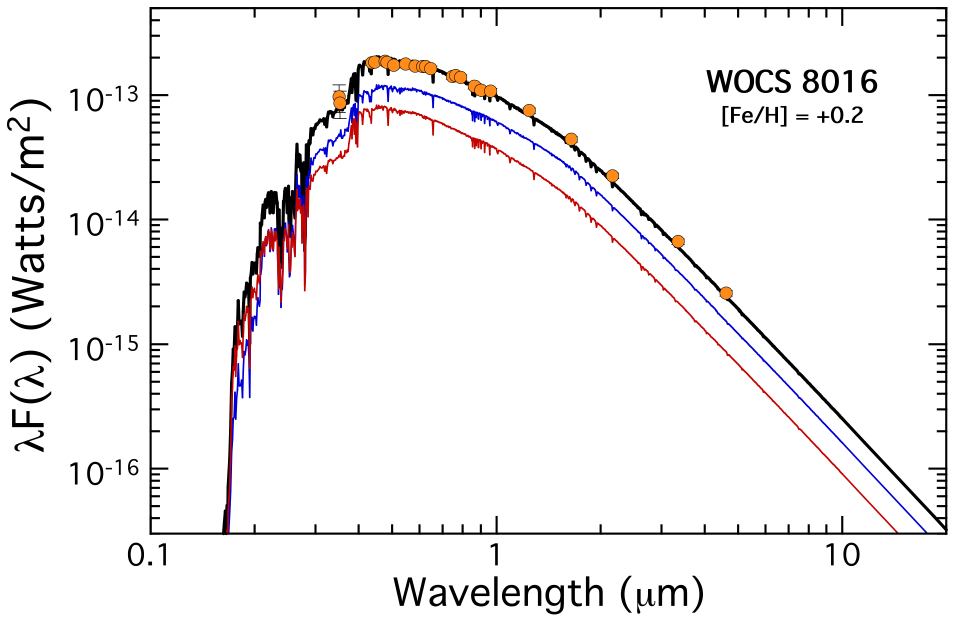}
\includegraphics[width=0.48\textwidth]{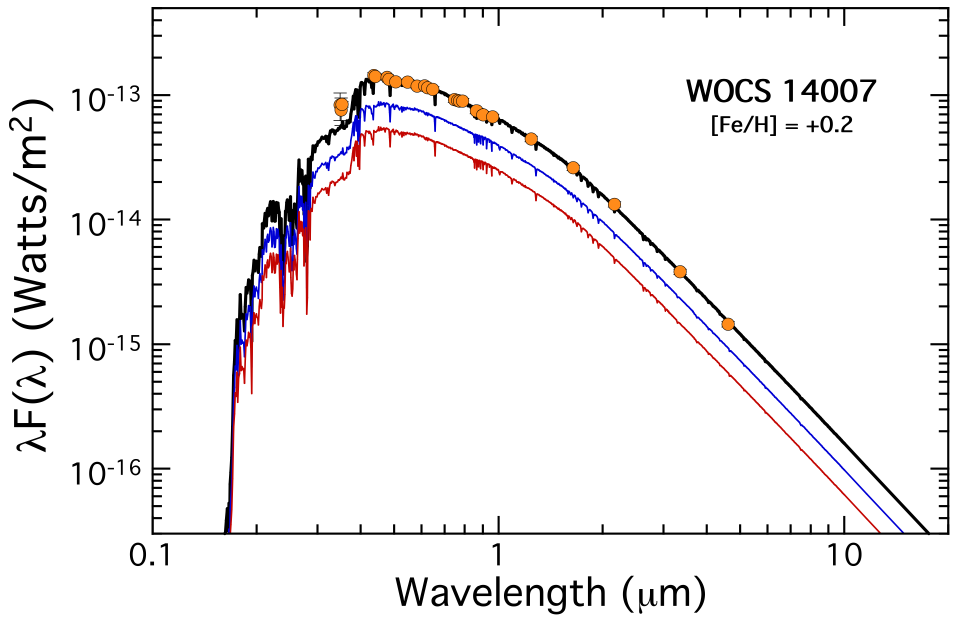}\\
\includegraphics[width=0.48\textwidth]{figures/SED_fit_17008_consistent_Fe+0.2.pdf}
\includegraphics[width=0.48\textwidth]{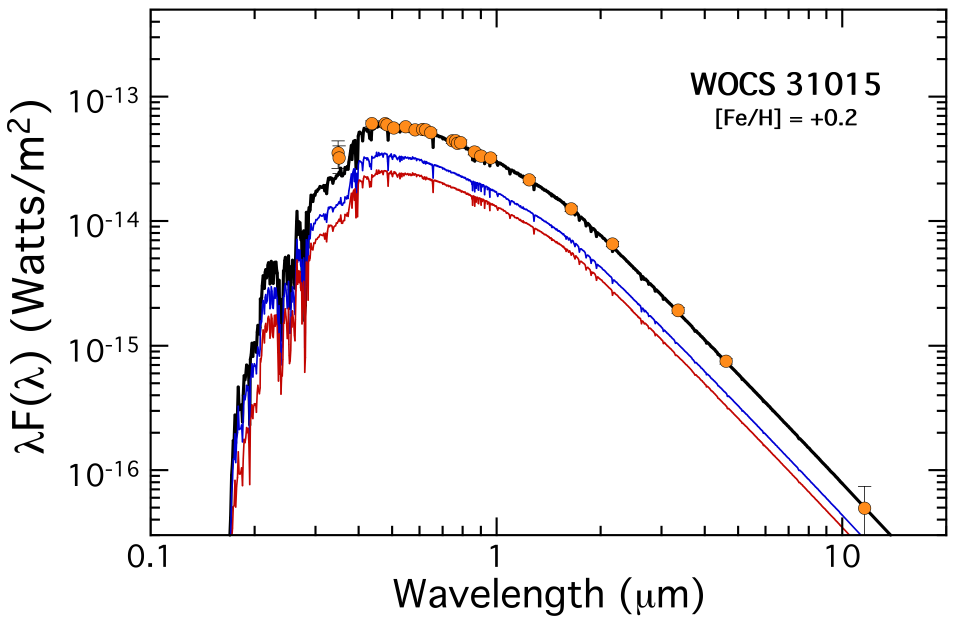}
\caption{We performed a joint SED fit for six binary systems, assuming a common age, distance, and extinction ($A_V$). The models were computed for a metallicity of [Fe/H] = 0.20. Figure~\ref{fig:age_Fe} illustrates how the derived age depends on metallicity. In the figure, blue and red lines represent the best-fit models for the individual components of each binary, while the black line corresponds to the combined (composite) SED. Observational data are shown as orange circles with associated error bars. }
\label{fig:seds}
\end{figure*}

The individual SEDs and their best-fitting models are shown in Fig.~\ref{fig:seds}. We find the overall quality of the fits quite satisfying.  The stellar and cluster parameters derived from these fits are summarized in Table~\ref{tab:ngc7789_sedresults}. 

\begin{figure}
\centering
\includegraphics[width=0.99\columnwidth]{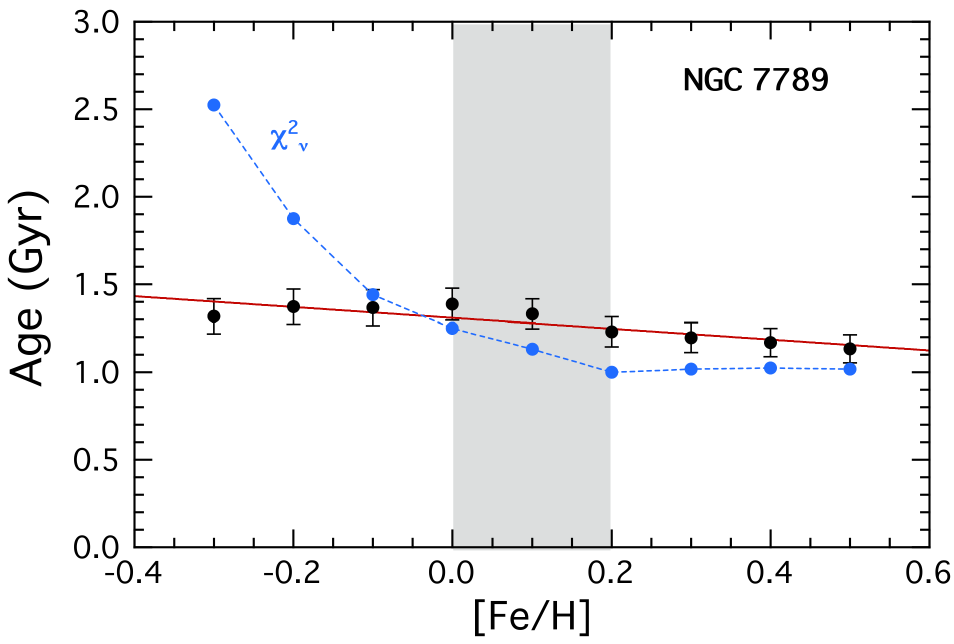}
\caption{Age of NGC~7789 vs.~the assumed metallicity [Fe/H]. The same type of SED fits used to produce Table \ref{tab:ngc7789_sedresults} and Fig.~\ref{fig:seds} was carried out for additional metallicities. We find a linear relation between the age of NGC~7789 and metallicity given in Sec.~\ref{sec:SEDs}. The $\chi^2$ value at each metallicity tested is also plotted to provide an idea of which metallicities are acceptable, i.e., for [Fe/H] $\gtrsim 0$.}
\label{fig:age_Fe}
\end{figure}

In order to quantify the sensitivity of the joint SED solution to the adopted chemical composition, we repeated the full six–binary fit across a grid of metallicities spanning $-0.3 \leq {\rm [Fe/H]} \leq +0.5$. For each metallicity point, the cluster age was independently re-derived from its posterior age distribution. The results are shown in Fig.~\ref{fig:age_Fe}. In particular, this grid explicitly includes the solar-metallicity case, which is found to follow the same overall age–metallicity trend.
Following the methodology applied in our earlier cluster studies 
\citep[see]{Yakut2025a,Yakut2025b}, the resulting age–metallicity relation can be described by a linear regression of the form

\begin{equation}
{\rm Age}({\rm Gyr}) = 1.26 - 0.31\,({\rm [Fe/H]} - 0.20),
\end{equation}
or equivalently,
\begin{equation}
{\rm Age}({\rm Gyr}) = 1.31 - 0.31\,{\rm [Fe/H]}.
\end{equation}

These expressions, plotted in Fig.~\ref{fig:age_Fe}, show a mild but systematic 
decrease in the inferred age with increasing metallicity for ${\rm [Fe/H]} > 0$.  
The minimum of the global $\chi^{2}$ curve occurs near ${\rm [Fe/H]} \simeq +0.20$, 
which we adopt as the preferred chemical composition of NGC\,7789.  
At this metallicity, the joint SED analysis yields a cluster age of 
$1.26 \pm 0.09$~Gyr (see also Fig.~\ref{fig:age_Fe}). The error bar cited here is
statistical only.  There is also another systematic component due to the uncertainty in metallicity of $\sim$$\pm 0.1$ dex, which amounts to about 0.03 Gyr.

As mentioned above, the stellar parameters derived from the SED fitting for a fixed metallicity [Fe/H] = +0.2 --- masses, radii, effective temperatures, luminosities, and orbital inclinations for all twelve stars --- together with the cluster-level parameters are listed in Table~\ref{tab:ngc7789_sedresults}. The good agreement between the SED-based temperatures and radii and those inferred from the RV+LC analysis  (for two of the binaries), as well as with the positions of the stars in the evolutionary diagrams (see Sec.~\ref{sec:results}), demonstrates that the joint SED approach provides a powerful and largely independent check on the fundamental parameters of NGC~7789.

\begin{table*}
\centering
\caption{Derived stellar parameters for the twelve stars in our six NGC~7789 binaries from the joint SED fit. System-level components are labelled with uppercase IDs (primary = a, secondary = b). Cluster parameters from the global solution appear at the bottom. Results correspond to the adopted metallicity [Fe/H] = 0.20.}
\label{tab:ngc7789_sedresults}
\begin{tabular}{lllllll}
\hline
WOCS   & ID & Mass                & Radius           & Temperature   & Luminosity      & Inclination \\
       &    & (${M_{\odot}}$)     & (${R_{\odot}}$)  & ($K$)         & (${L_{\odot}}$) & (${{^\circ}}$) \\
\hline
17028 & Aa & $1.888 \pm 0.030$ & $2.962 \pm 0.143$ & $6727 \pm 153$ & $16.22 \pm 1.08$ & $81.8 \pm 0.8$\\
      & Ab & $1.388 \pm 0.034$ & $1.478 \pm 0.055$ & $6553 \pm  75$ & $ 3.67 \pm 0.43$ & $81.8 \pm 0.8$\\
14014 & Ba & $1.779 \pm 0.032$ & $2.423 \pm 0.110$ & $6917 \pm 123$ & $12.16 \pm 0.96$ & $82.2 \pm 0.8$\\
      & Bb & $1.532 \pm 0.033$ & $1.745 \pm 0.064$ & $6802 \pm  64$ & $ 5.92 \pm 0.58$ & $82.2 \pm 0.8$\\
 8016 & Ca & $1.928 \pm 0.039$ & $3.252 \pm 0.169$ & $6600 \pm 130$ & $18.15 \pm 1.56$ & $73.9 \pm 2.8$\\
      & Cb & $1.767 \pm 0.038$ & $2.382 \pm 0.150$ & $6919 \pm 126$ & $11.77 \pm 1.19$ & $73.9 \pm 2.8$\\
14007 & Da & $1.790 \pm 0.043$ & $2.474 \pm 0.159$ & $6900 \pm 130$ & $12.56 \pm 1.38$ & $44.3 \pm 1.5$\\
      & Db & $1.628 \pm 0.057$ & $1.968 \pm 0.139$ & $6892 \pm  85$ & $ 7.97 \pm 1.30$ & $44.3 \pm 1.5$\\
17008 & Ea & $1.962 \pm 0.041$ & $3.531 \pm 0.186$ & $6591 \pm 155$ & $21.23 \pm 1.54$ & $54.5 \pm 1.2$\\
      & Eb & $1.441 \pm 0.044$ & $1.570 \pm 0.069$ & $6660 \pm  91$ & $ 4.43 \pm 0.60$ & $54.5 \pm 1.2$\\
31015 & Fa & $1.494 \pm 0.028$ & $1.669 \pm 0.049$ & $6749 \pm  59$ & $ 5.25 \pm 0.44$ & $64.1 \pm 1.7$\\
      & Fb & $1.405 \pm 0.028$ & $1.507 \pm 0.045$ & $6589 \pm  63$ & $ 3.89 \pm 0.37$ & $64.1 \pm 1.7$\\
\hline 
\multicolumn{3}{c}{{Age}} & \multicolumn{3}{c}{{$A_V$}} & {} \\
\multicolumn{3}{c}{(Myr)} & \multicolumn{3}{c}{(mag)} & {} \\
\hline
\multicolumn{3}{c} {$1257 \pm 90$} & \multicolumn{3}{c}{{$0.90 \pm 0.05$}} & {} \\
\hline
\hline
\end{tabular}

Notes: The distance prior was tightly constrained around the Gaia distance of $2080 \pm 142$ pc, and the metallicity was fixed at [Fe/H] = +0.2 (but see Figure~\ref{fig:age_Fe} for an exploration of the cluster metallicity.)
\end{table*}

\section{Astrometric Analysis}
\label{sec:astrometric}

\begin{figure}
\centering
\includegraphics[width=1.1\linewidth]{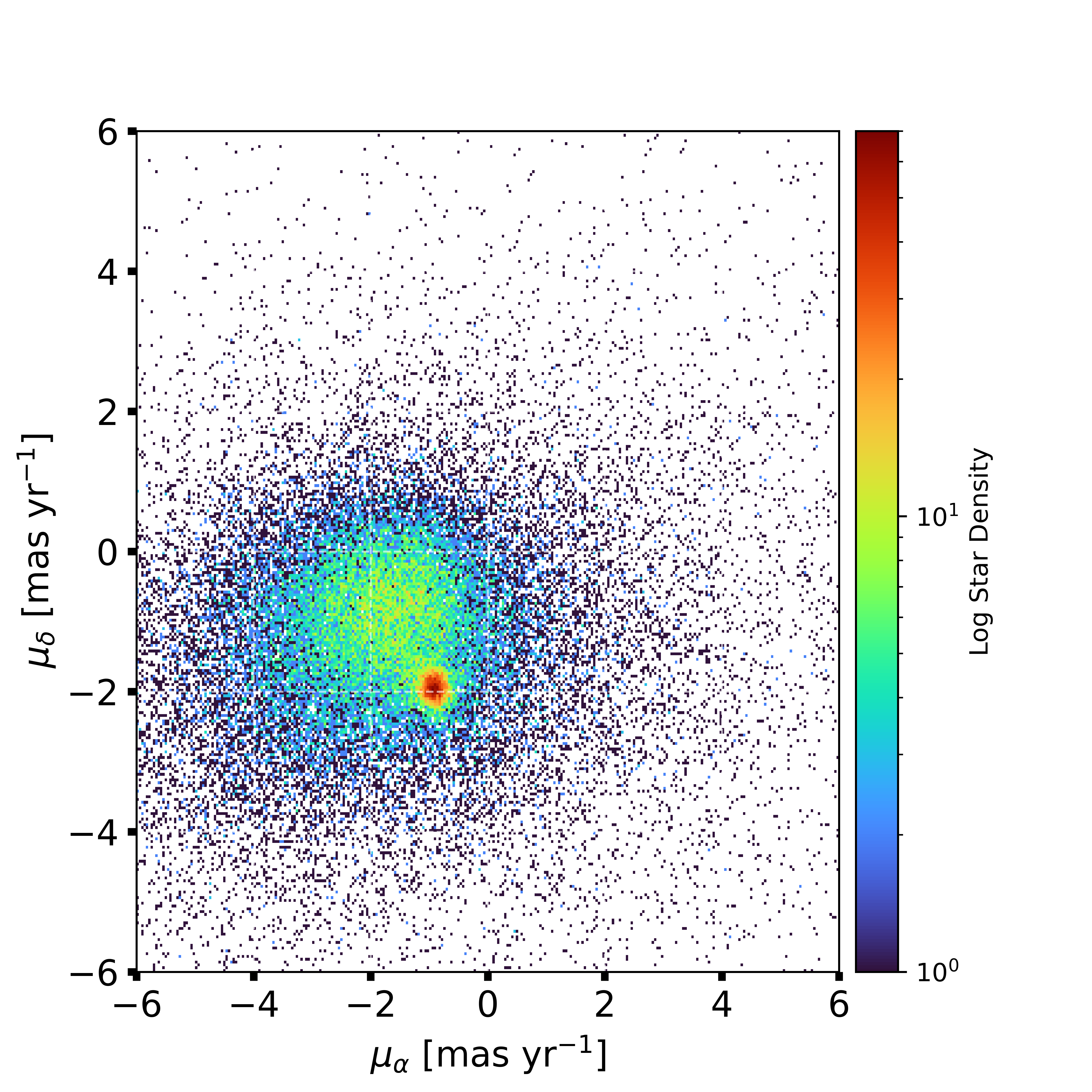} 
\includegraphics[width=0.85\linewidth]{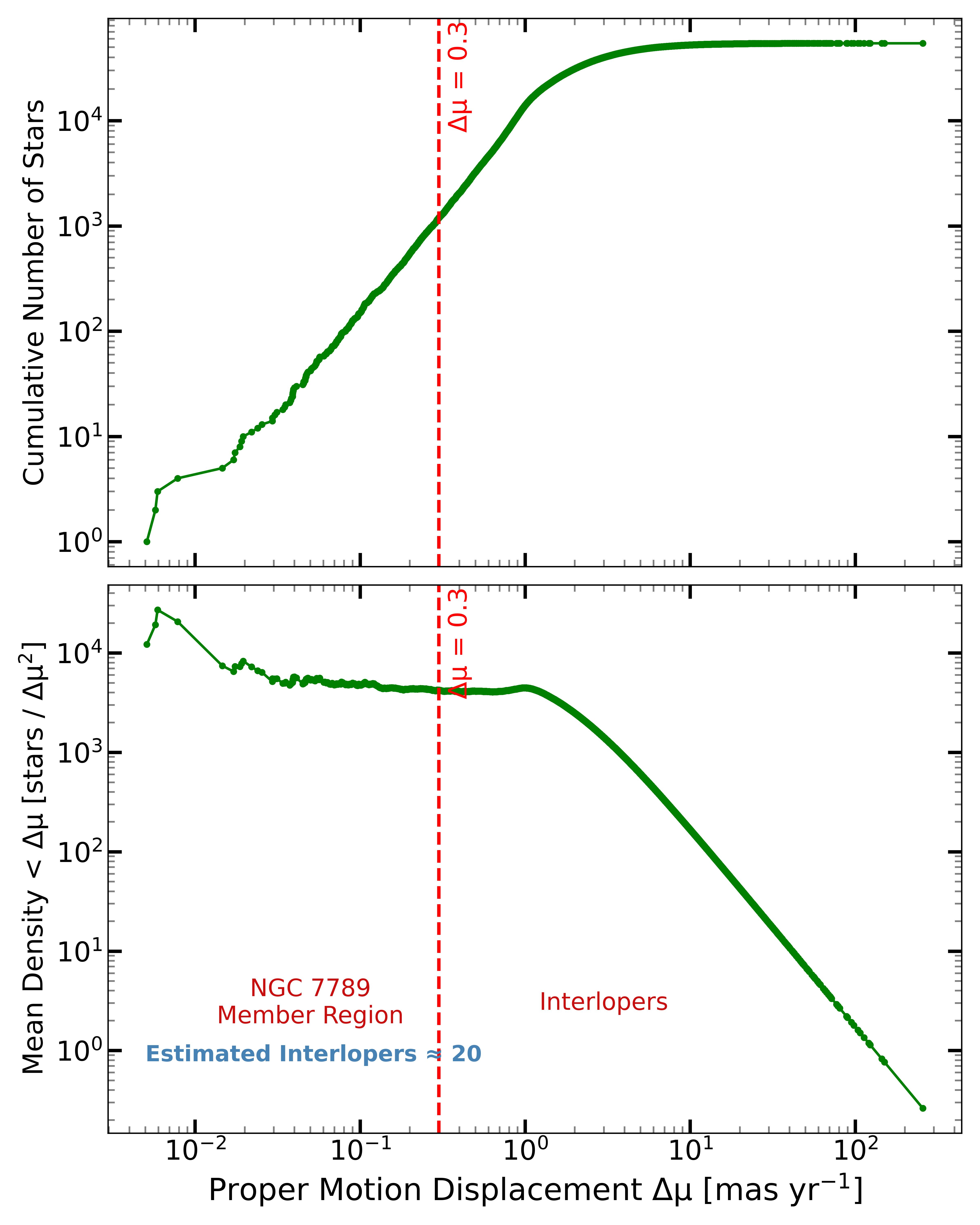} 
\caption{\textbf{Top:} Proper–motion density map of the NGC\,7789 field based on \textit{Gaia} DR3 data. The cluster population is clearly identified as the prominent red overdensity centred at 
$(\mu_{\alpha*},\mu_\delta) \simeq (-0.97,\,-1.89)$~mas\,yr$^{-1}$, corresponding to the mean cluster motion. \textbf{Middle:} Cumulative distribution of proper–motion displacements $\Delta\mu$ from the cluster centroid. The vertical dashed line at $\Delta\mu = 0.3$~mas\,yr$^{-1}$ marks the adopted membership boundary separating the dense cluster core from the surrounding field population. 
\textbf{Bottom:} Mean cumulative density (number of stars per $\Delta\mu^{2}$) as a function of $\Delta\mu$, showing a rapid decline in source density beyond the adopted cutoff and confirming the effectiveness of the membership selection.}
\label{fig:pm_density_ngc7789}
\end{figure}

The astrometric properties of NGC\,7789 were examined using \textit{Gaia} DR3 data in order to establish a consistent and contamination-free membership list prior to the combined RV, LC, and SED modelling. Following the procedure adopted in our earlier binary-based age studies \citep{Yakut2025a, Yakut2025b}, we first applied quality filters on the Gaia measurements---parallax uncertainties, RUWE values, and photometric excess factors---to ensure a reliable astrometric sample. Candidate members were then identified using a multidimensional selection in proper-motion space, parallax, and position on the Gaia CMD. This resulted in a high–probability member list of 1296 stars. The cluster’s proper-motion centroid was determined to be $(\mu_{\alpha*},\mu_\delta) = (-0.972 \pm 0.177,\,-1.894 \pm 0.166)$~mas\,yr$^{-1}$, with a mean parallax of $\varpi = 0.464 \pm 0.004$~mas, corresponding to a distance of \(2.08 \pm 0.14\)~kpc, where the quoted uncertainty reflects the dispersion among member distances (standard deviation) rather than just the formal error on the mean. After defining a high-probability membership list based on Gaia~DR3 astrometry, we adopted the individual geometric distances ($r_{\rm geo}$) from the Bailer-Jones catalogue \citep{BailerJones2021} for each confirmed member star.  These geometric distances are derived from a Bayesian framework that explicitly accounts for the Gaia~DR3 parallax zero-point offset and other known astrometric systematics, and therefore no additional parallax zero-point correction was applied in this work. These star-by-star distances incorporate Gaia parallax zero-point corrections and other known systematics within a Bayesian framework. The cluster distance and its uncertainty were then derived from the statistical distribution of the $r_{\rm geo}$ values of the selected members, rather than from a direct inversion of the mean parallax. The principal astrometric and global properties adopted for NGC\,7789 are summarised in Table~\ref{tab:ngc7789_astrometry}. The resulting distribution of high-probability members in proper-motion space is presented in Figure~\ref{fig:pm_density_ngc7789}.

\begin{table}
\centering
\caption{Summary of the astrometric, kinematic, and global properties adopted for NGC\,7789.}
\label{tab:ngc7789_astrometry}
\begin{tabular}{lc}
\hline
Parameter & Value \\
\hline

Cluster centre RA (J2000) & $23^{\rm h}57^{\rm m}24^{\rm s}$ \\
Cluster centre Dec (J2000) & $+56^{\circ}42'30''$ \\

$\mu_{\alpha*}$ (mas\,yr$^{-1}$) & $-0.972 \pm 0.177$ \\
$\mu_{\delta}$ (mas\,yr$^{-1}$) & $-1.894 \pm 0.166$ \\

Mean parallax $\varpi$ (mas) & $0.464 \pm 0.004$ \\
Distance (kpc) & $2.08 \pm 0.14$ \\

Mean radial velocity $\langle V_{\rm r} \rangle$ (km\,s$^{-1}$) & $-54.9 \pm 0.5$ \\

Binary-based age (Gyr) & $1.26 \pm 0.09$ \\

Number of Gaia DR3 members & 1296 \\

\hline
\end{tabular}
\end{table}

To refine the membership list further, we employed a Gaussian-mixture-based probabilistic model, assigning membership likelihoods to all stars within a radius of $20'$ around the cluster centre. Only stars with membership probabilities exceeding $90\%$ were retained. This ensures that the binaries used in the subsequent SED and RV+LC analysis are genuine cluster members sharing a common kinematic origin. The astrometric selection also helps suppress field contamination along this crowded Galactic-plane sightline, where proper-motion overlaps between field and cluster stars are substantial. The final membership list implies an expected field contamination of only about 20 stars out of 1296.

The main-sequence turnoff (MSTO), located near $G \approx 13.8$~mag and 
$(BP-RP) \approx 1.05$\,mag, is fully consistent with the binary-based cluster age of 
$1.26 \pm 0.09$~Gyr derived in this work, and lies within the range inferred by previous 
isochrone-based studies of NGC\,7789. 
Notably, most of binaries included in our analysis are located in the evolved portion 
of the CMD, making them especially valuable as independent age diagnostics, since their 
radii and masses can be determined dynamically.

Finally, the astrometric analysis provides the global cluster framework adopted
as external priors in our combined SED and RV+LC modelling. The cluster distance of $2.08 \pm 0.14$~kpc, derived from Gaia~DR3 astrometry, was used as a tight prior in the SED analysis, thereby defining the common geometric scale for all six binary systems.
The line-of-sight extinction was allowed to vary within a broad prior range ($A_V = 0.74$--$0.98$~mag), and the joint SED fitting converged on $A_V = 0.90 \pm 0.05$~mag, fully consistent with the adopted prior and with the cluster colour--magnitude diagram morphology.
The agreement between the astrometric framework and the binary-based modelling demonstrates that the inferred stellar and cluster parameters are mutually consistent within a single, coherent solution for NGC\,7789.

\begin{figure}
\centering
\includegraphics[width=1.0\linewidth ]{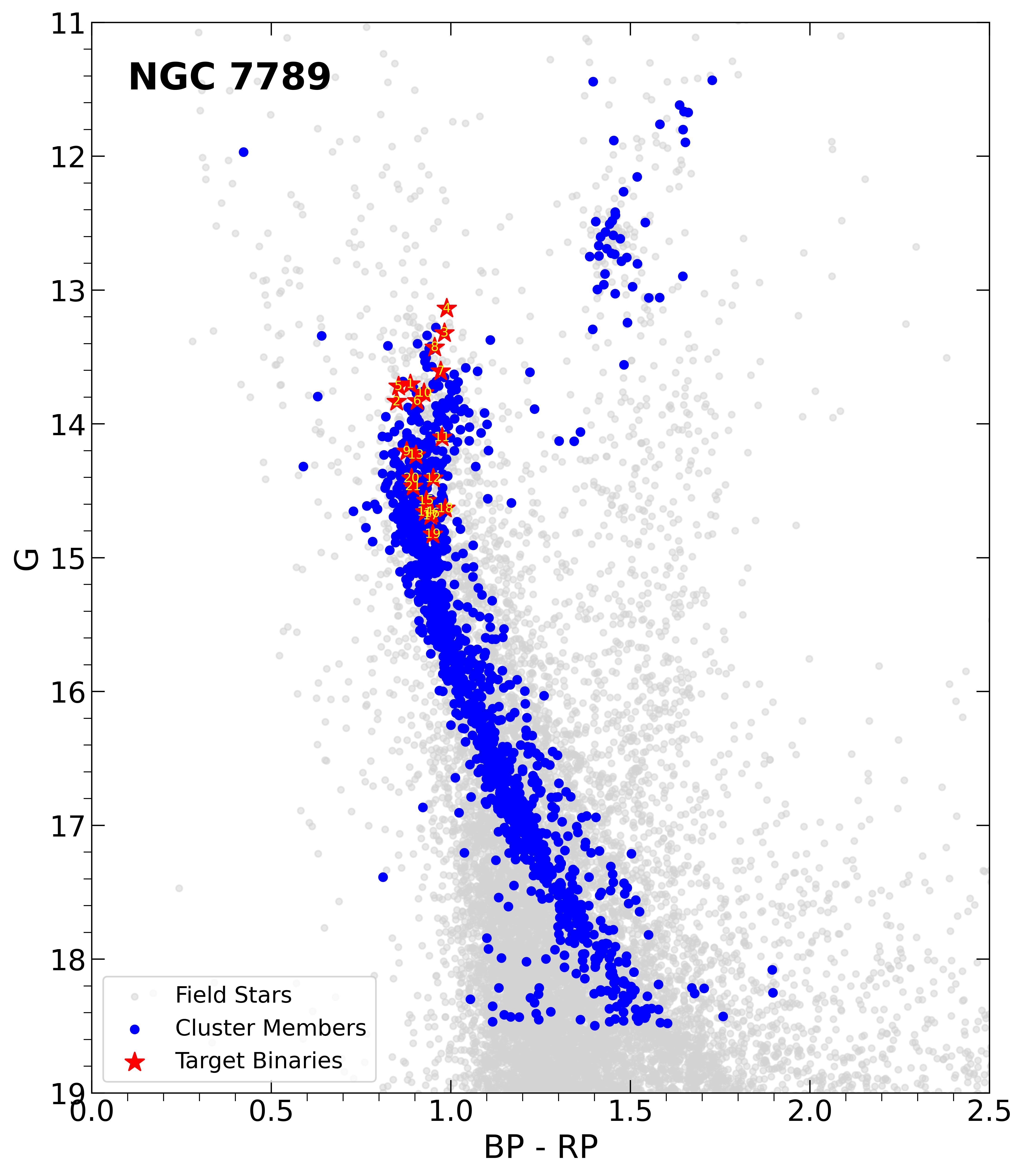} 
\caption{Colour–magnitude diagram of NGC\,7789 based on \textit{Gaia} DR3 photometry. 
The vertical axis shows the $G$ magnitude. 
All sources within a 20-arcmin radius of the cluster centre are shown in gray, 
while high-probability members selected through astrometric and photometric criteria are highlighted in blue. The twelve stars of the six target binaries are marked as red star symbols with overplotted numeric labels corresponding to their IDs. Most of these binary components are evolved stars located close to the cluster main-sequence turnoff, where age sensitivity is maximised.}
\label{fig:gaiaDR3HR}
\end{figure}

\section{Results and Conclusion}
\label{sec:results}

We have carried out a comprehensive determination of the fundamental parameters of the open cluster NGC\,7789 by jointly analysing six well‑characterised binary systems using radial velocity (RV) measurements, TESS photometry, and broadband spectral energy distributions.  
This work extends the binary‑based SED+LC+RV methodology introduced by our team in recent cluster studies, where it successfully yielded high‑precision, model‑independent ages for both the old benchmark cluster NGC\,188 \citep{Yakut2025a} and the intermediate‑age system NGC\,2506 \citep{Yakut2025b}.  
Applied to NGC\,7789, this framework provides an alternative and complementary route to conventional isochrone‑only techniques by leveraging the absolute masses, radii, and luminosities of evolved detached binaries.

With the SEDs for all six targets, and incorporating the latest dynamical constraints, our joint modelling converges on a cluster age of
$1.26 \pm 0.09$~Gyr.  
This value lies at the younger end of, but remains consistent with, the canonical $1.2$–$1.6$~Gyr range inferred from modern Gaia‑based isochrone analyses 
\citep[e.g.,][]{Dias2021,Balan2024,Ferreira2025}.  
The age–metallicity trend obtained from the six binaries shows a weak but systematic decline in age with increasing [Fe/H], and the global $\chi^2$ minimum indicates a preferred metallicity of ${\rm [Fe/H]} \approx +0.20$.  
This value lies within 2-3$\sigma$ of the uncertainties of published spectroscopic measurements  
\citep{Tautvaisiene2005,Overbeek2015,Carrera2013}  
and produces excellent agreement between the predicted and observed radii and effective temperatures of all binary components.

The joint analysis further demonstrates that the stellar parameters inferred
from the binaries are fully consistent with the astrometric properties of the
cluster. The Gaia~DR3 distance of $2082 \pm 142$~pc, adopted as an external prior in the
SED modelling, is entirely compatible with the luminosities and radii inferred
for all six binary systems. Within this framework, the SED fitting converges on a line-of-sight extinction of $A_V = 0.90 \pm 0.05$~mag, which lies comfortably within the adopted prior
range and is consistent with literature reddening estimates as well as with the observed cluster sequence in the Gaia colour--magnitude diagram. This agreement confirms that the binary-based modelling and the astrometric constraints together define a coherent and self-consistent global solution for NGC\,7789.

The SED fits yield mutually consistent stellar parameters for all six systems.
Only one source, WOCS\,17028, shows a residual discrepancy at the $\sim2\sigma$ level
between the stellar masses inferred from the radial-velocity orbit and those implied
by the SED-based modelling. Given its broad photometric wavelength coverage, such a mismatch is plausibly explained by a modest level of unrecognised third‑light contamination rather than a deficiency in the binary evolution modelling. Crucially, excluding this system from the joint analysis leaves the derived cluster parameters well within the uncertainties, demonstrating the robustness and redundancy of the multi‑binary approach.

It is also worth mentioning that for four of the six systems, reliable stellar masses were obtained without relying on inclination angles derived from eclipses or ellipsoidal variability. In these cases, the combination of double-lined spectroscopic orbits with the joint SED constraints is sufficient to determine both the component masses and the inclination angles. This is possible because the RVs provide the mass ratios which are crucial to separating the contribution from the two stars in a given composite SED.  This underscores the applicability of the method to non-eclipsing binaries.

Overall, NGC\,7789 becomes the third open cluster for which the SED+LC+RV framework 
delivers high-precision and physically well-constrained stellar and cluster parameters—frequently with more reliable age constraints than those available from isochrone fitting alone.  
The strong internal consistency between the binary-derived age, reddening, and adopted 
metallicity underscores the diagnostic power of evolved detached binaries in 
intermediate-age clusters.  
Future high-resolution spectroscopy and additional long-period binary characterisation 
within NGC\,7789 will further refine the cluster metallicity and may reduce the remaining 
statistical uncertainties in the age determination.

\begin{figure}[h]
\centering
\includegraphics[width=0.45\textwidth]{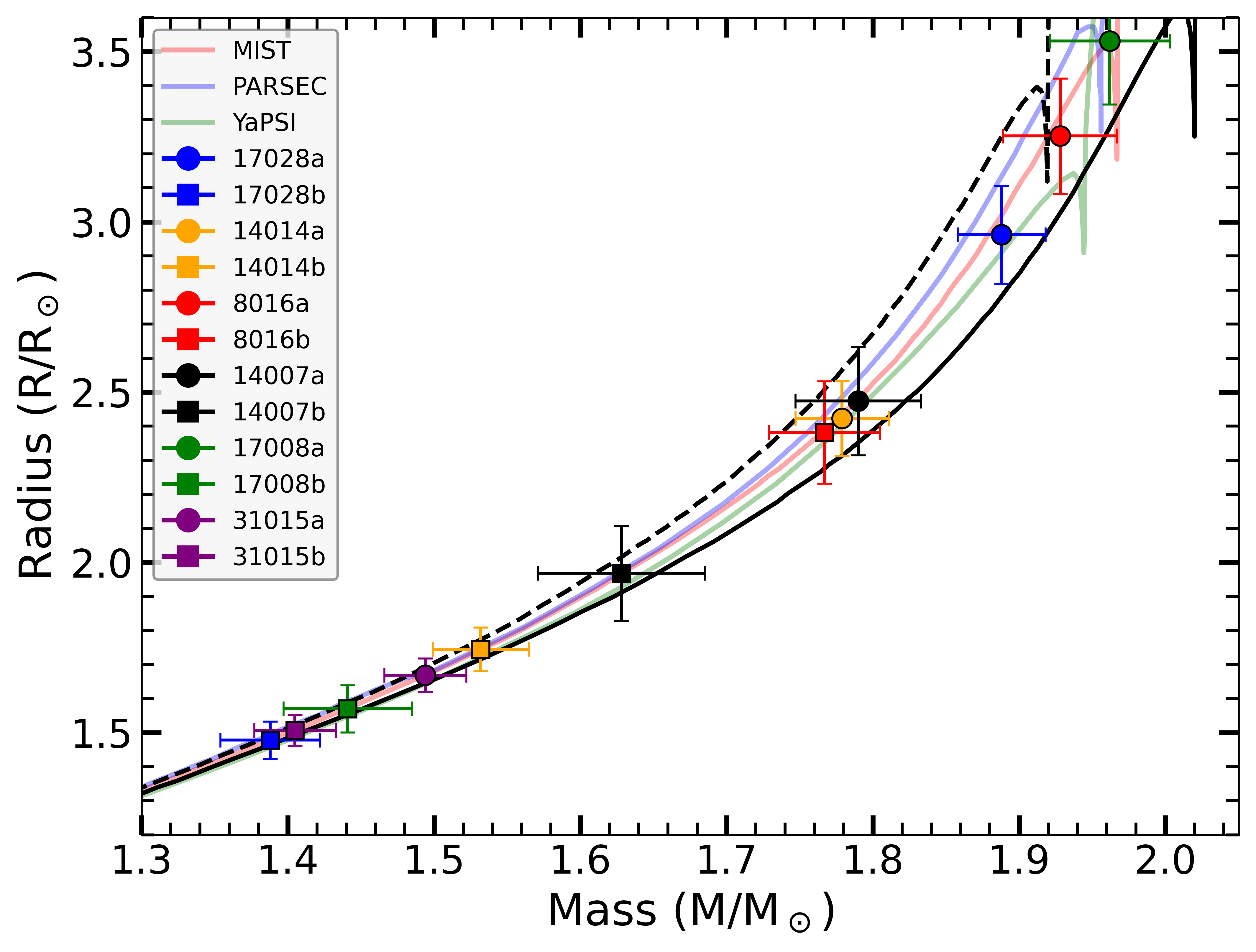}
\includegraphics[width=0.45\textwidth]{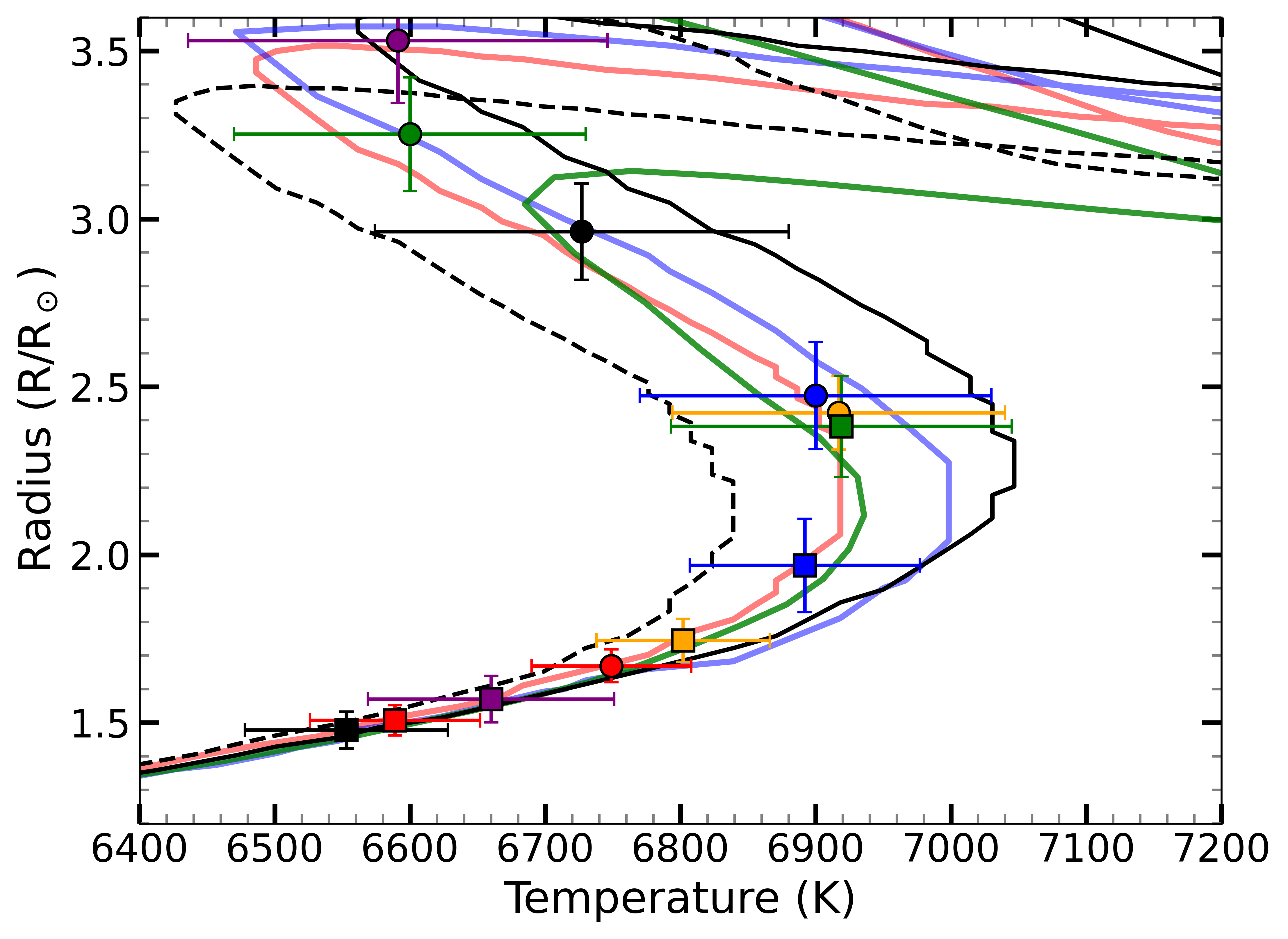}
\includegraphics[width=0.48\textwidth]{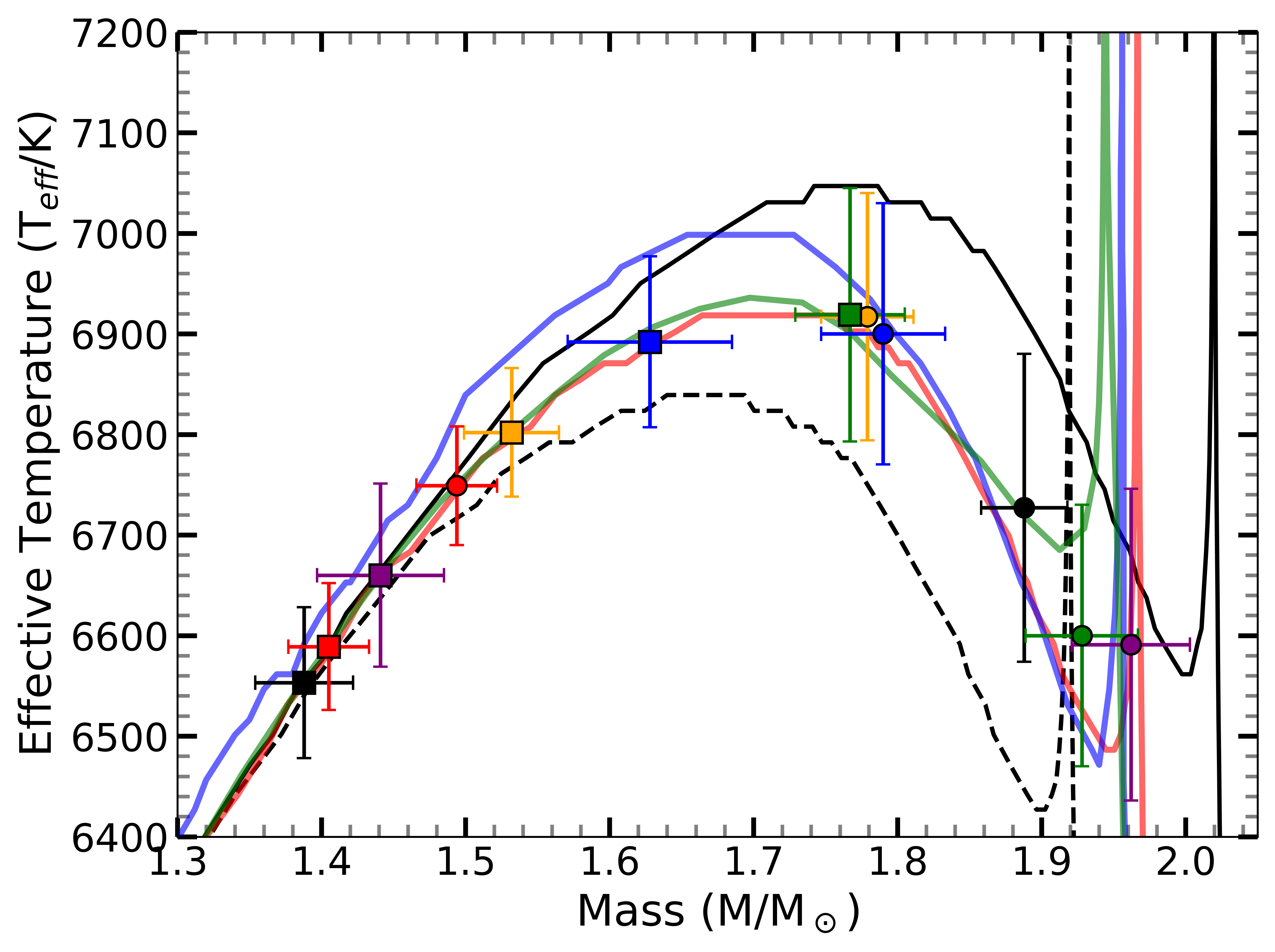} 
\caption{Radius–Mass, Radius–Temperature, and Temperature–Mass diagrams for the six binary systems analysed in NGC\,7789. Each binary component is shown as a different color marker. Stellar evolution isochrones from MIST (red solid), PARSEC (blue dashed), and YaPSI/Y² (green dashed) for an age of $1.26 \pm 0.09$~Gyr are overplotted.  
To illustrate the age uncertainty range, additional MIST isochrones at ages of $1.32$~Gyr (black dashed) and $1.14$~Gyr (black solid) are included. The parameters of all six binaries are listed in Table~\ref{tab:ngc7789_sedresults}.}
\label{fig:iso}
\end{figure}

\section*{Acknowledgements}
We thank the anonymous referee for their careful reading of the manuscript and for several constructive comments that helped to improve the clarity and robustness of this work.  This work has made use of data from the European Space Agency (ESA) \textit{Gaia} mission (\url{https://www.cosmos.esa.int/gaia}), processed by the Gaia Data Processing and Analysis Consortium (DPAC; \url{https://www.cosmos.esa.int/web/gaia/dpac/consortium}). It also makes use of observations obtained with the Transiting Exoplanet Survey Satellite (\textit{TESS}) mission, funded by the NASA Explorer Program and publicly available through the Mikulski Archive for Space Telescopes (MAST). This research has made use of the SIMBAD and VizieR catalogues operated at the Centre de Données astronomiques de Strasbourg (CDS), and of NASA’s Astrophysics Data System (ADS) bibliographic services. This study was supported by the Scientific and Technological Research Council of T\"urkiye (TÜBİTAK, grant numbers 112T766, 117F188 and 2219). Numerical computations were performed in part using the High Performance and Grid Computing Center (TRUBA resources) provided by TÜBİTAK ULAKBİM. KY gratefully acknowledges the Master and Fellows of Churchill College, University of Cambridge, for the award of an Overseas Research Fellowship.

\section*{Data Availability}

The TESS observations used in the analysis of binary light curves are available online to the public through the Mikulski Space Telescope Archive (MAST). If desired, the data used in this study can be obtained from the MAST servers or requested from the authors.



\bibliographystyle{mnras}
\bibliography{ngc7789} 



\bsp	
\label{lastpage}
\end{document}